\documentclass [11pt]{article}

\usepackage{amsmath,amssymb,epsfig,cite,color,verbatim}
\usepackage{graphics,float}
\usepackage{overpic} 
\usepackage{mathrsfs}
\usepackage{slashed,bbold}
 \usepackage{multicol, array}
\usepackage{cancel}
\usepackage{soul}

\usepackage[normalem]{ulem}

\numberwithin{equation}{section}

\title{Quantum phase transition and superselection sectors in two-color one-flavor matrix QCD at weak coupling
}
\date{\today}
\author{Nirmalendu Acharyya$^1$\footnote{nirmalendu@iitbbs.ac.in}, Arkajyoti Bandyopadhyay$^1$\footnote{s22ph09003@iitbbs.ac.in}, Parthsarathi~Kondapure$^{2}$\footnote{kparthsarath@iisc.ac.in} and Sachindeo~Vaidya$^2$\footnote{vaidya@iisc.ac.in} \\
${}^1${\small School of Basic Sciences, Indian Institute of Technology Bhubaneswar, Jatni, Khurda, Odisha 752050, India}\\
${}^2${\small Centre for High Energy Physics,  Indian Institute of Science, Bengaluru, 560012, India}\\
}
\begin{document}

\maketitle
\begin{abstract}

 We numerically investigate the weak-coupling (small $g$) regime of the matrix model of $SU(2)$ gauge theory coupled to one flavor of massless fundamental quarks. The theory possesses a global  Pauli-G\"ursey $SU(2)_B$  symmetry, which organizes physical states of spin $J$ into sectors labelled by its representation $B$. We find localization-delocalization quantum phase transitions (QPTs) in multiple $(B,J)$ sectors at sector-dependent critical couplings $g_\ast^{BJ}$. Direct evaluation of the localization probability reveals that the ground states are predominantly localized near $A_i=0$ for $g<g_\ast^{BJ}$, but become almost entirely expelled from this region for $g>g_*^{BJ}$. The QPTs in different sectors exhibit distinct features, and we establish that quark-glue interactions are essential for these transitions. The transitions are singular:  the ground state expectation value of $\text{Tr}(A_iA_i)$ remains finite for $g< g_\ast^{BJ}$, whereas it diverges as $g$ approaches $g_\ast^{BJ}$ from above.  
We show that the two limiting ground states on either side of $g_\ast^{BJ}$ belong to distinct superselection sectors. Furthermore,  as $g$ approaches $g_\ast^{BJ}$ from above, the ground state  
localizes on a submanifold (of codimension 2) of the gauge configuration space,  
with the spin and color of its glue components becoming locked.

\end{abstract}
\section{Introduction}

A recent study of the matrix model of $SU(2)$ gauge theory coupled to one flavor of adjoint quark identified a very interesting quantum phase transition (QPT) in the weak coupling (strength denoted by $g$) regime\cite{Acharyya:2026uhx}. The QPT is robustly established by discontinuities in ground state expectation values of several observables.  
Particularly intriguing, however, is the behavior of the ground state wavefunction in gauge configuration space, which makes this transition truly special. Evidences, though indirect, from several gauge invariant observables suggested that the transition involves a dramatic change in the localization properties of the ground state wavefunction:
below a critical coupling $g_\ast$, the ground state remains predominatly localized near the origin of the gauge configuration space, whereas for $g>g_\ast$, the wavefunction becomes delocalized over configurations with large gauge fields.

These observations raise several natural questions.  First, is this localization-delocalization phase transition a peculiarity of the adjoint matter, or is this a more general phenomenon that can occur in theories with fundamental quarks (or in pure gauge theory)? This distinction is physically significant because fundamental and adjoint fermions have different gauge and global-symmetry structures. In particular, for $SU(2)$ gauge theory, the pseudoreality of the fundamental representation gives rise to an enlarged Pauli--G\"ursey symmetry, which relates quark and antiquark degrees of freedom and plays a crucial role in the  internal composition of physical states. Thus, the existence of the phase transition for adjoint matter does not necessarily imply its persistence in presence of the fundamental quarks as well. Second, besides the indirect evidences in \cite{Acharyya:2026uhx}, can the localization-delocalization phenomenon be probed directly, and what is its microscopic origin? In particular, what role does the quark-glue interaction  play in driving this transition?  Last but not the least, what is the nature of the critical behavior, and does the transition entail a deeper change in the structure of the physical Hilbert space?

To address these questions, we investigate the matrix model of two-color gauge theory coupled to one flavor of fundamental quark.  This model has global $SU(2)_B$ Pauli-G\"ursey symmetry, and the physical color-singlet states can be neatly arranged according to their spin $J$ and in representation of $SU(2)_B$, labelled by the Casimir $B(B+1)$. We investigate the possibility of the localization-delocalization transition in  the lowest-energy  spin-0 and spin-1 states belonging to different $B$-sectors.

An important aspect of our investigation is the direct characterization of the ground state wavefunction in gauge configuration space. The structure of the configuration space is fundamental to understanding gauge dynamics \cite{Asorey:1986vr, Asorey:1996klg, Pause:1998by, Asorey:1998wm, Aguado:2000aw}. However, probing it directly in the full field theory is highly challenging \cite{Greensite:1979yn, Feuchter:2004mk}, and the matrix model provides a tractable setting in which this structure can be studied explicitly. Here, we  characterize the localization properties of the wavefunction by computing its probability fraction within a specified region of gauge configuration space. This provides a direct probe of the configuration-space structure of the ground state and allows identifying the QPT  without relying solely on indirect spectral signatures, as in \cite{Acharyya:2026uhx}.

We find that the  localization-delocalization transition occurs in the multiple spin-0 and spin-1 sectors belonging to different representations of $SU(2)_B$.   The localization property of the transition remains the same for all these sector:  the ground state is completely localized near $A_i=0$ for $g<g_\ast^{BJ}$, while it is almost entirely expelled from that region for $g>g_\ast^{BJ}$. However, the detailed nature and microscopic manifestations of the transition in these sectors are not identical.  In some sectors, the transition is accompanied by a discontinuous change in the quark content and/or in the spin-fraction carried by the quarks, whereas other sectors are devoid of such changes.  Thus, although the localization transition occurs across several symmetry sectors, the associated restructuring of the internal fermionic composition is sector-dependent.   
Furthermore, we find that several $(B,J)$ sectors in which the quark--glue interaction (and quark--glue entanglement) is absent do not exhibit the transition. 
 This observation provides insight into the mechanism underlying the phenomenon and indicates an essential role for the quark-glue interaction.   
 At very small $g$, this interaction only provides a perturbative correction to the dynamics of the decoupled quark and glue and the ground state (lightest energy eigenstate) remains nearly separable, with predominant support on configurations with small gauge fields. As the coupling grows to $g_\ast^{B, J}$, the attractive quark-glue interaction substantially lowers the energy of the state with support on large gauge configurations,  favoring the delocalization.

Finally, we investigate the critical behavior in  different $(B,J)$ sectors by examining the nature of the corresponding ground states at the critical couplings $g_\ast^{BJ}$. For each sector, we construct the limiting ground state wavefunctions by approaching the critical coupling from the two sides of the transition. To characterize these limiting states, we compute the expectation value $\Phi \equiv \langle\text{Tr} A_i A_i\rangle_{gs}$, which quantifies the extent of the corresponding wavefunction in the gauge configuration space.  Interestingly, we find that for all $(B,J)$ sectors with the transition, $\Phi$ remains finite for the limiting state obtained from the localized phase, whereas it diverges for the state approached from the delocalized phase. Furthermore, we find that in the $(B,J)=(0,0)$ sector, this divergence follows a power law behavior, $\Phi\sim(g-g_\ast^{00})^{-2}$, in the vicinity of $g_\ast^{00}$ in the delocalized phase.   
The quantity $\sqrt{\Phi}$ defines a characteristic ``length scale"  in configuration space and its divergence has profound and far-reaching consequences. We argue that the divergence in $\sqrt{\Phi}$ originates from the wavefunction acquiring support on the matrix configuration with identical singular values and this in turn leads to spin-color locking \cite{Pandey:2016hat}.  More remarkably, we show that the limiting ground state wavefunctions obtained from opposite sides of the transition belong to different superselection sectors. These distinct superselection sectors appear only at the critical point: the ground states at generic $g$ in the bulk of either phase belong to the same superselection sector.  Thus, the localization--delocalization transition is accompanied by a singular change in the configuration-space extent of the ground state and, more fundamentally, in the superselection structure of the physical Hilbert space.

The paper is organized as follows. In Sec.~\ref{sec_2}, we introduce the matrix model of one-flavor fundamental QC$_2$D and discuss its symmetries and physical Hilbert space. We focus our discussion of the spin-0 and spin-1 states, which can belong to $B=0,1,2$ and $B=0,1$ sectors, respectively.  In Sec.~\ref{sec_results}, we present the results of our findings. The global ground state of the theory belongs to the $(B,J)=(0,0)$ sector. In section \ref{B0J0_sec}, we show that the lowest energy state of this sector undergoes the localization-delocalization transition and depict the properties of the two phases. Further, we show similar transition occurs in the $(B,J)=(1,1)$ and $(B,J)=(0,1)$ sectors as well. When we compare the properties of the phases in these three sectors, we identify how each one is distinct in character from the other. Contrastingly, in the $(B,J)=(1,0)$ and $(B,J)=(2,0)$ sectors, we show that the ground states do not undergo any sharp transition. In both these sector, the quark-glue interaction vanishes, and we identify that this interaction plays a key role in the phase transition. Interestingly, the ground state in $(B,J)=(1,0)$ sector is inherently entangled and shows a relatively fast crossover, which is absent in the completely separable states of  $(B,J)=(2,0)$ sector.  In Sec.~\ref{critical_phono},  we analyze the limiting states at the critical coupling and establish their superselection properties. Using a cubic invariant, we show that one of these limiting ground state localizes on the surface of configurations with identical singular values and argue that it corresponds to spin-color locking.  We conclude in Sec.~\ref{results_sec} with a discussion of our results.

\section{One-flavor matrix QC$_2$D} \label{sec_2}
The construction of the matrix model involves pulling back the Maurer-Carter one-form of $SU(2)$ to a spatial three-sphere $S^3$ of radius $R$ \cite{Balachandran:2014iya, Balachandran:2014voa}. The gauge (glue) degrees of freedom are described by a  real matrix $M_{ia}$ (with the spatial index $i=1,2,3$ and the color index $a=1,2,3$), or equivalently by $A_i \equiv M_{ia} T^a$, where $T^a$'s are the generators of $SU(2)$ in the fundamental representation.  The phase space of the glue dynamics is generated by $M_{ia}$ and its conjugate momenta $P_{ia}=-i \frac{\partial}{\partial M_{ia}}$. The chromoelectric and chromomagnetic fields are given by $\Pi_{i}= P_{ia}T^a$ and $\mathcal{B}_{i}= - R^{-1} A_i - \frac{i}{2} \epsilon_{ijk} [Aj, A_k]$. The Hamiltonian for the pure glue dynamics is 
\begin{eqnarray}
H_{YM}= \text{Tr}\Big[\frac{g^2}{R^3} \Pi_i \Pi_i + \frac{R^3}{g^2}\mathcal{B}_i \mathcal{B}_i \Big]. 
\end{eqnarray}

The quarks in the matrix model are introduced as time-dependent Grassmann-valued matrices \cite{Pandey:2016hat,Pandey:2019dbp}. In the quantum mechanical model, choosing the Weyl representation, the quarks can be represented by $\psi= (b_{\alpha A},\, d_{\alpha A}^\dagger)^T$, where $b_{\alpha A}$ and  $d^\dagger_{\alpha A}$ satisfy the standard anti-commuting algebra.  These quarks interact with the glue via minimal coupling and the interaction Hamiltonian is \cite{Acharyya:2024pqj}:  $H_{int}= R^3 (b^\dagger_{\alpha A} \sigma^i_{\alpha \beta} T^a_{AB} b_{\beta B}-d_{\alpha A} \sigma^i_{\alpha \beta} T^a_{AB} d^\dagger_{\beta B})M_{ia}$. The quarks can in general be massive: $H_m=mR^3(b_{\alpha A}^\dagger d_{\alpha A}^\dagger + h.c)$. In this work, we will focus only on the chiral case (i.e. $m=0$). However, it is straightforward to adapt the discussion for the massive case.

It is convenient to work with dimensionless variables and so we rescale $A_i \to g R^{-1} A_i$, $\Pi_{i} \to g^{-1} R \Pi_i$, $b_{\alpha A} \to R^{-\frac{3}{2}} b_{\alpha A}$, and $d_{\alpha A} \to R^{-\frac{3}{2}} d_{\alpha A}$. Under this rescaling, the Hamiltonian massless quarks becomes 
\begin{eqnarray}
H= \frac{1}{R}  \Big[\text{Tr}(\Pi_i \Pi_i) +V_{YM}+ g (b^\dagger_{\alpha A} \sigma^i_{\alpha \beta} T^a_{AB} b_{\beta B}-d_{\alpha A} \sigma^i_{\alpha \beta} T^a_{AB} d^\dagger_{\beta B})M_{ia} 
\Big]. 
\end{eqnarray}
where the Yang-Mills potential above is given by
\begin{eqnarray}
V_{YM} \equiv \text{Tr}\Big( A_i A_i +i g \epsilon_{ijk} [A_i, A_j] A_k- \frac{g^2}{2} [A_i, A_j]^2\Big) 
\end{eqnarray}

The extrema of the Yang-Mills potential $V_{YM}$ are static classical solutions of the pure YM theory, and as we will argue that  they play a significant role in determining the quantum dynamics. This potential is minimized when the gauge configurations satisfy $[A_i, A_j]=i g \epsilon_{ijk}A_k$. This leads to two solutions: trivial minimum at $A_i=0$, and non-trivial minima at $A_i =\frac{1}{2g} \sigma_i$, with $V_{YM}=0$ for both. There is another extrema at $A_i =\frac{1}{4g} \sigma_i$, which corresponds to a saddle point with $V_{YM}=\frac{3}{32 g^2}$. Computing $V''_{YM}$ yields the frequencies of the small oscillations $\omega$ around the minima. Around the trivial minimum, $\omega=1$ (ninefold degenerate) and around the non-trivial minimum, $\{\omega\}=\{2,1,0\}$ with degeneracy$=\{5,1,3\}$. The three zero modes around the  non-trivial minimum correspond to the gauge transformations. 

In the quantum theory, the modes around the trivial minima have the zero-point energy $9\times\frac{1}{2}=\frac{9}{2}$, while those around the non-trivial one have $\Big(5\times\frac{2}{2}+ 1 \times \frac{1}{2}+3 \times \frac{0}{2}\Big)=\frac{11}{2}$. Thus, there is a significant  gap between the energies of the lightest modes around the trivial and non-trivial minima. Though $V_{YM}$ has the structure of a double well, it is not symmetric in the quantum theory. 

When $g$ is very small, the barrier at the saddle point is very high and the two minima are widely separated. In such a situation, to the leading order, the low energy effective Hamiltonian is that of a harmonic oscillator around each well, with  the cubic, the quartic, and the interaction terms acting as small perturbations. The ground  state is approximately the lightest state  of the oscillator around the trivial minimum and has energy $9/2 - O(g^2)$. Because the lightest state of the oscillator around the non-trivial minimum has significantly higher energy, the  perturbations are too weak to generate a significant tunnelling amplitude between the two wells. Consequently, the ground state for very small $g$ practically remains localized in the well around the trivial minimum.

The situation changes as $g$ increases because the barrier height and separation between the minima decrease and the interaction strength increases. For sufficiently large $g$, the attractive quark-glue interaction can significantly lower the effective potential around the non-trivial minimum.  In such a case, it might be energetically favourable for the wavefunction to get delocalized to configurations with large $A_i$.

This implies that in between these two situations, the system undergoes a quantum phase transition at $g_\ast$ where the ground state's localization property changes drastically. This yields a localized (near $A_i=0$) phase for $g<g_\ast$ and a  delocalized one for $g>g_\ast$.

\section{Results}\label{sec_results}
The colorless states belong to the physical Hilbert space $\mathcal{H}_{phys}$  and these states can be arranged accoroding to their spin $J$ and in irreps of the $SU(2)_B$ (Pauli-G\"ursey), labelled by $B$ (see Appendix \ref{App_symm_Hphys}). The physical states with fixed $B$ and $J$ forms sectors in the $\mathcal{H}_{phys}$.  
In the following, we diagonalize  the Hamiltonian numerically using the variational techniques and probe the properties of the ground state.  In particular, we show that there is are QPTs in the weak coupling regimes for different $(B, J)$ sectors. Further, by direct estimation of the localization property of these ground state  in the gauge configuration space, we establish that the QPTs correspond to the localization-delocalization transition discussed above. We will demonstrate that  the quark-glue interaction is the key driving factor for this transition to occur.

For the numerical simulations, we expand the bosonic wavefunctions in the basis of harmonic oscillator energy eigenstates and truncate the expansions at a maximum number of oscillator quanta $N_b$ \cite{Acharyya:2024pqj, Acharyya:2026uhx}. We progressively increase $N_b$ until convergence is achieved in the energy eigenvalues. In general, the numerical estimates of the energies converge well with $N_b \sim 20$: the results for  $N_b>20$ are all practically indistinguishable. Near the phase transitions at $g_\ast$, the convergence is slow (fails completely exactly at $g_\ast$). This is not an artefact of the numerics, rather there is a signature of the critical phenomena. We use the  finite size scaling in $N_b$ to study the region in the vicinity of the critical point. 

\begin{figure}[t]
\begin{center}
\includegraphics[width=16cm]{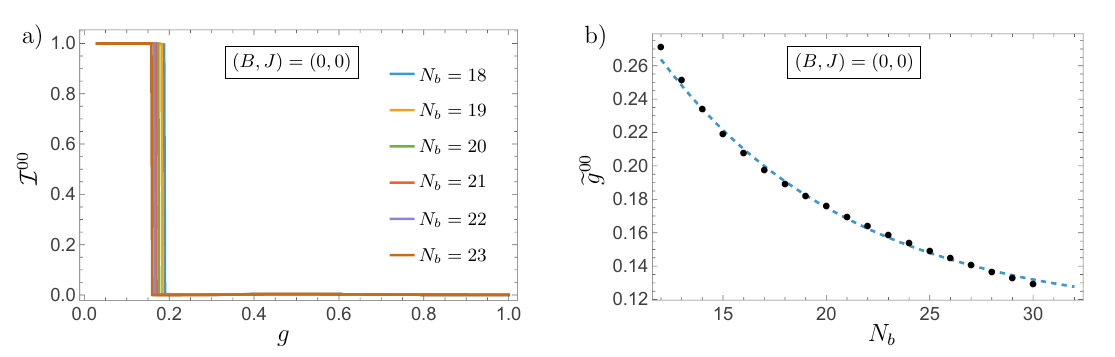}
\caption{a) The localization measure $\mathcal{I}^{00}$ vs $g$ for different $N_b$. b) $\widetilde{g}^{00}$ vs $N_b$. The dashed line corresponds to the fit to (\ref{fit_g}). }\label{Fig_1}
\end{center}
\end{figure}

\subsection{$(B,J)=(0,0)$ sector}\label{B0J0_sec}
The global ground state  of the theory belongs to the $(B,J)=(0,0)$ sector for all values of the coupling $g$.  We numerically compute the ground state energy $E_0(g)$ and the ground state wavefunction $|\Psi_{\mathrm{gs}}\rangle$ in the weak coupling regime $0 \leq g \lesssim 1$. To quantify the localization properties of the ground state in the potential well around the trivial minimum at $M_{ia}=0$ (equivalently, $A_i=0$), we define a 9D rectangular box $\Gamma=\{M_{ia}: \, -\frac{1}{2g}\leq M_{ia} \leq \frac{1}{2g}\}$.  Since the saddle point is at $A_{i}=\frac{1}{4g}\sigma_{i}$, this region  encloses the potential well around $M_{ia}=0$. Using the ground state wavefunction $|\Psi_{\mathrm{gs}}\rangle$, we define the \textit{localization measure} 
\begin{eqnarray}
\mathcal{I}^{00}(g)\equiv \int_\Gamma\rho_{_\mathcal{G}}(M) \,d\mu_M, \quad\quad \rho_{_\mathcal{G}}(M)\equiv \langle\Psi_{\mathrm{gs}}|\Psi_{\mathrm{gs}}\rangle_{_F} \quad\quad d\mu_M=\prod_{i,a=1}^3 dM_{ia}
\end{eqnarray}
where  $\langle \,\cdot| \cdot\, \rangle_{_F}$  denotes the inner
product in the fermionic Hilbert space, and $\rho_{_\mathcal{G}}(M)$ is the probability density in the gauge configuration space.

\begin{figure}[h]
\begin{center}
\includegraphics[width=16cm]{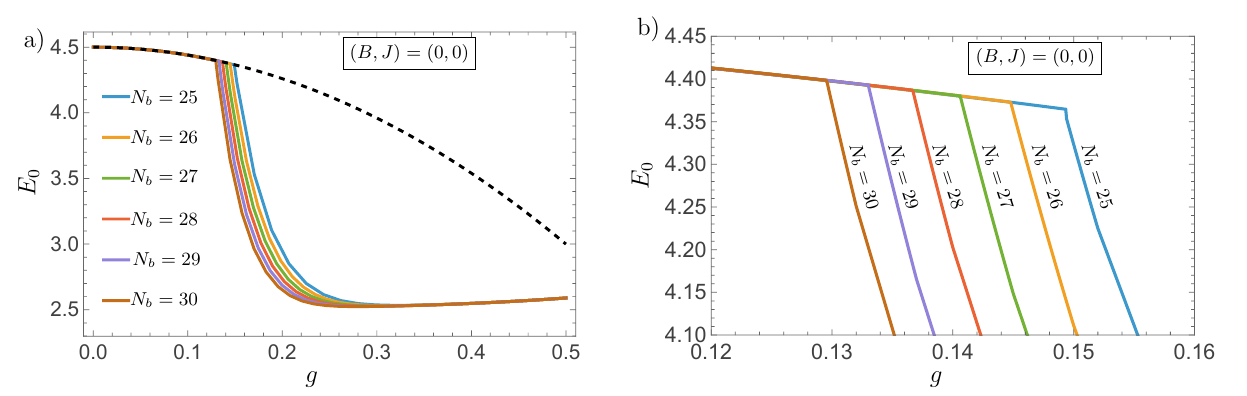}
\caption{a) $E_0$ vs $g$ for different $N_b$ in the $(B,J)=(0,0)$ sector.  The black dashed line shows the estimates from second order perturbation theory.  b) Zooming in the region around $\widetilde{g}^{00}(N_b)$. }\label{Fig_2}
\end{center}
\end{figure}

Figure~\ref{Fig_1}a shows $\mathcal{I}^{00}$ as a function of $g$ for different values of the bosonic cutoff $N_b$.  
Interestingly, we find that for every $N_b$, there exist a coupling $\widetilde{g}^{00}(N_b)$ such that 
\begin{eqnarray}
\mathcal{I}^{00}(g,N_b) \left\{ \begin{array}{ll}
\simeq  1, \quad & \text{for } g< \widetilde{g}^{00}(N_b), \\ [0.75 ex]
\ll 1,  \quad & \text{for } g> \widetilde{g}^{00}(N_b). 
\end{array}\right.
\end{eqnarray}
In particular, above $\widetilde{g}^{00}(N_b)$, the probability of finding the ground state inside $\Gamma$ becomes negligibly small. This indicates a substantial redistribution of the ground state probability away from the potential well surrounding the trivial minimum.

We next examine the dependence of $\widetilde{g}^{00}(N_b)$ on the bosonic cutoff (Fig.~\ref{Fig_1}b). Similar to the adjoint quark case studied in \cite{Acharyya:2026uhx}, the numerical data for $\widetilde{g}^{00}(N_b)$ can well described by
\begin{eqnarray}
\widetilde{g}^{00}(N_b) \simeq g_\ast^{00}  - c e^{- k N_b}, 
\label{fit_g}
\end{eqnarray}
where $g_\ast^{00}$,  $c$, and $ k$ are given in Table~\ref{Table_1}.  
This behavior suggests that  in the $N_b \to \infty$ limit, $\widetilde{g}^{00}(N_b)$ converges to $g_\ast^{00}$ and there is a dramatic change in the localization property of the ground state at $g_\ast^{00}$: 
\begin{eqnarray}
\lim_{N_b \to \infty} \mathcal{I}^{00}(g, N_b) \left\{ \begin{array}{ll}
\simeq  1, \quad & \text{for } g<g_\ast^{00}, \\ [0.75 ex]
\ll 1,  \quad & \text{for } g>g_\ast^{00}. 
\end{array}\right.
\end{eqnarray}
 For $g<g_\ast^{00}$, the ground state is almost entirely localized in the region $\Gamma$, while for $g>g_\ast^{00}$, its probability within this region becomes
negligible. The ground state for $g>g_\ast^{00}$ is therefore expelled from the region surrounding the trivial minimum and becomes delocalized over gauge configurations outside $\Gamma$. This sharp change in the localization properties signals that there is a QPT at $g_\ast^{00}$, separating a localized phase for $g<g_\ast^{00}$ from a delocalized phase for $g>g_\ast^{00}$. 
\begin{figure}[t]
\begin{center}
\includegraphics[width=16cm]{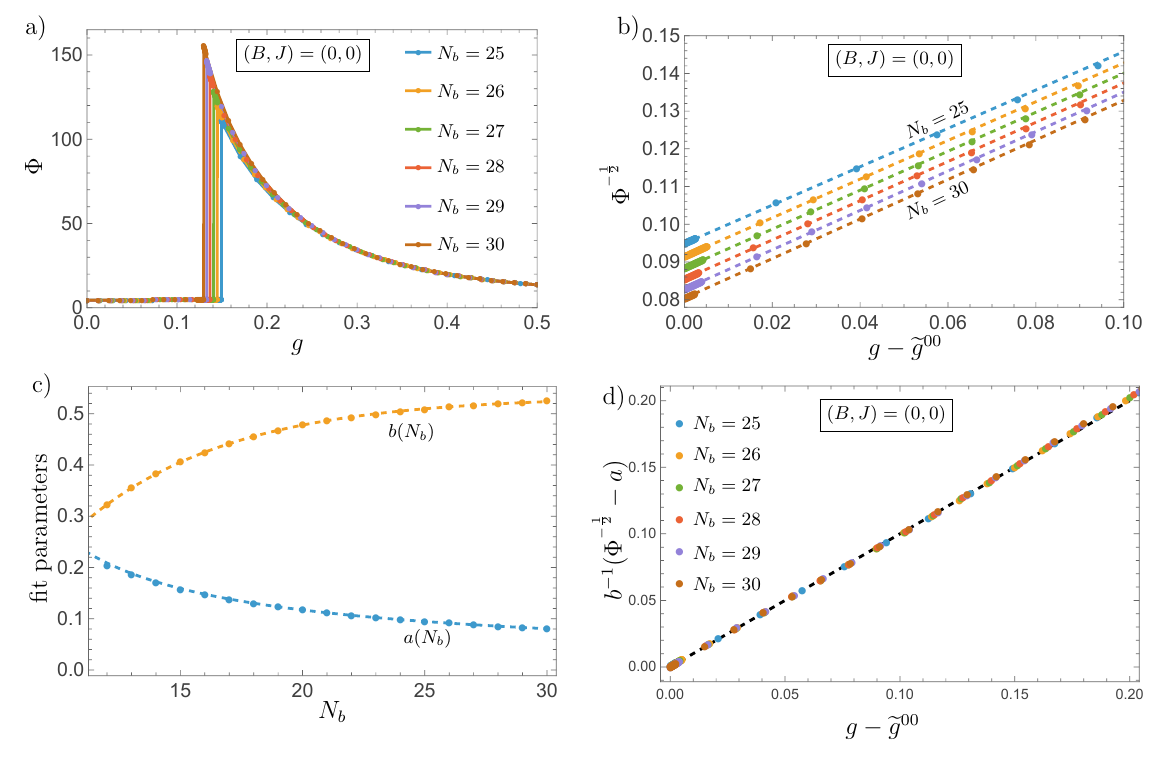}
\caption{a) $\Phi$ vs $g$ for different $N_b$ in $(B,J)=(0,0)$ sector. b) $\Phi^{^{\frac{1}{2}}}$ vs $g-\widetilde{g}^{00}$ for $g>\widetilde{g}^{00}$ with different $N_b$. The dashed lines corresponds to the fit (\ref{fit_invsqrtphi}). c) The fit parameters $a(N_b)$ and $b(N_b)$ as functions of $N_b$. The dashed lines denote the fits (\ref{fit_parameters}).  d) Log-plot of $b^{-1}(\Phi^{^{-\frac{1}{2}}} -a)$ vs $(g-\widetilde{g}^{00})$ for different $N_b$. The black dashed line represents $y=x$, on which the data for all $N_b$ collapse to.   }\label{Fig_3}
\end{center}
\end{figure}

In this weak coupling regime, the ground state energy $E_0$ (in units of $R^{-1}$) is shown in Fig.~\ref{Fig_2}a.  As evident from the figure,  $E_0(g)$ develops a kink for every $N_b$, whose location coincides with $\widetilde{g}^{00}(N_b)$. In the vicinity of this kink, on the side $g > \widetilde{g}^{00}(N_b)$,  the numerical data exhibits a breakdown in convergence, as illustrated in Fig.~\ref{Fig_2}b. The convergence is gradually restored as the coupling is increased further, away from the kink.

At very small $g$, the ground state can be well approximated by the lightest state of a 9D harmonic oscillator with $\mathcal{O}(g^2)$ perturbative corrections from the cubic, quartic, and interaction terms.  It is straightforward to see that to the leading order, second order perturbation theory yield 
\begin{eqnarray}
E_{0}(g) \simeq \frac{9}{2} - 6 g^2. 
\end{eqnarray}
Indeed, for $g<\widetilde{g}^{00}(N_b)$, the numerical estimates of the ground state energy has excellent match with this perturbative result (shown as dashed line in Fig.~\ref{Fig_2}a).

For all $g > \widetilde{g}^{00}(N_b)$, however, $E_0(g)$ deviates significantly from the perturbative estimate, indicating the dominance of the non-perturbative effects in this regime.  Moreover, there is a sharp drop in the ground state energy  at $g=\widetilde{g}^{00}(N_b)+ \epsilon$, with the slope becoming progressively steeper as $N_b$ increases.  These indicates that in the converged limit with $N_b \to \infty$, the ground state energy becomes non-analytic at $g_\ast^{00}$ and hence provides an independent energetic signature of the QPT at $g_\ast^{00}$.

For the observable $\mathcal{O}\equiv \frac{2}{R^2}\text{Tr}(A_i A_i)$, we define its ground state expectation value  as $\Phi(g)= \langle \mathcal{O}\rangle_{gs}$. The dependance of $\Phi$ (in units of $R^{-2}$) on the coupling $g$ is shown in Fig.~\ref{Fig_3}a. For $g<\widetilde{g}^{00}(N_b)$, $\Phi(g)$ remains small and nearly constant. 
However, for ach value of $N_b$, $\Phi(g)$ exhibits a sharp peak at $\widetilde{g}^{00}(N_b)$ whose height increases progressively with $N_b$. Consequently, $\Phi(g)$ fails
to converge at $\widetilde{g}^{00}(N_b)$. For $g \gg \widetilde{g}^{00}$, the convergence of $\Phi(g)$ is restored, and its value remains finite. 

The failure of convergence at $\widetilde{g}^{00}(N_b)$ is not a numerical artefact; rather, it is the signal of the underlying physical phase transition.  
 Interestingly, we find that $\Phi^{^{-\frac{1}{2}}} $ grows linearly with positive $(g-\widetilde{g}^{00})$ for every $N_b$ (Fig.~\ref{Fig_3}b).  This
behavior is well described by 
\begin{eqnarray}
\Phi^{^{-\frac{1}{2}}} (g, N_b) \simeq a + b\Big(g-\widetilde{g}^{00}(N_b)\Big), \quad\quad a(N_b), b(N_b) \in \mathbb{R}, \label{fit_invsqrtphi}
\end{eqnarray}
where $a(N_b)$ and $b(N_b)$ are, respectively, the intercept and slope obtained from linear fits (the dashed lines in Fig.~\ref{Fig_3}b). The dependence of these parameters on the bosonic cutoff $N_b$ is shown in Fig.~\ref{Fig_3}c , which can described well by
\begin{eqnarray}
&& a(N_b) \simeq \frac{1}{A + B N_b^{\gamma}}, \quad\quad  b(N_b) \simeq C- D \, e^{-\beta N_b}.\label{fit_parameters}
\end{eqnarray}
  The parameters $A$, $B$, $\gamma$ are given in Table~\ref{Table_2}, and  $C\simeq 0.534\pm 0.001$, $D\simeq 1.53\pm 0.03$ and $\beta \simeq 0.165 \pm 0.002$. 
Finally, Fig.~\ref{Fig_3}d shows that the quantity $b^{-1}(\Phi^{^{-\frac{1}{2}}} -a)$ plotted against $g-\widetilde{g}^{00}(N_b)$,  is independent of $N_b$, demonstrating a data collapse.

\begin{table}
\begin{center}
{\small
\begin{tabular}{|c|c|c|c|c|c|c|c|}
\hline & &&&\\ 
 & & $(B,J)=(0,0)$  & $(B,J)=(1,1)$  & $(B,J)=(0,1)$ \\
  & &  sector &  sector &  sector \\[0.25ex]
 \hline &&&&\\
Parameters & $g_\ast^{BJ}$ & $0.109\pm 0.003$ & $0.118\pm 0.003$ & $0.122 \pm 0.002$\\[0.5ex]
for  (\ref{fit_g})& $c$ & $0.55\pm 0.03$ & $0.63 \pm 0.03$& $ 0.82 \pm 0.03$ \\  [0.5ex]
 and (\ref{fit_g_1101})& $k$ &  $0.107\pm 0.006$ & $0.099 \pm 0.005$ & $0.112\pm 0.003$\\ [0.5ex] \hline
\end{tabular}}
\caption{The parameters for the fits in  (\ref{fit_g}) and  (\ref{fit_g_1101}).}\label{Table_1}
\end{center}
\end{table}

\begin{table}
\begin{center}
{\small
\begin{tabular}{|c|c|c|c|c|c|c|c|}
\hline & &&&\\ 
 & & $(B,J)=(0,0)$  & $(B,J)=(1,1)$  & $(B,J)=(0,1)$ \\
  & &  sector &  sector &  sector \\[0.25ex]
 \hline &&&&\\
Parameters& $A$ & $- (5.98\pm 0.01) $ & $-(7.38 \pm 0.01)$ & $-(9.02 \pm 0.02) $ \\  [0.5ex]
for & $B$  & $2.52 \pm 0.01$  & $3.18 \pm 0.01$    & $3.39 \pm 0.07$ \\ [0.5ex]
 (\ref{fit_parameters}) \& (\ref{fit_phi_B1J1_B0J1})& $\gamma$ & $0.585\pm 0.002$ & $0.515 \pm 0.001$ & $0.521 \pm 0.006$ \\ [0.5ex] \cline{1-5} 
\end{tabular}}
\caption{The parameters for the fits in (\ref{fit_parameters}) and  (\ref{fit_phi_B1J1_B0J1}). }\label{Table_2}
\end{center}
\end{table}

As $N_b\to\infty$, the intercept vanishes while the slope saturates to a non-zero constant:
\begin{eqnarray}
\lim_{N_b \to \infty} a(N_b) \simeq 0, \quad\quad \lim_{N_b \to \infty}b(N_b) \simeq 0.534 \pm 0.0007. 
\end{eqnarray}
\begin{figure}
\begin{center}
\includegraphics[width=16cm]{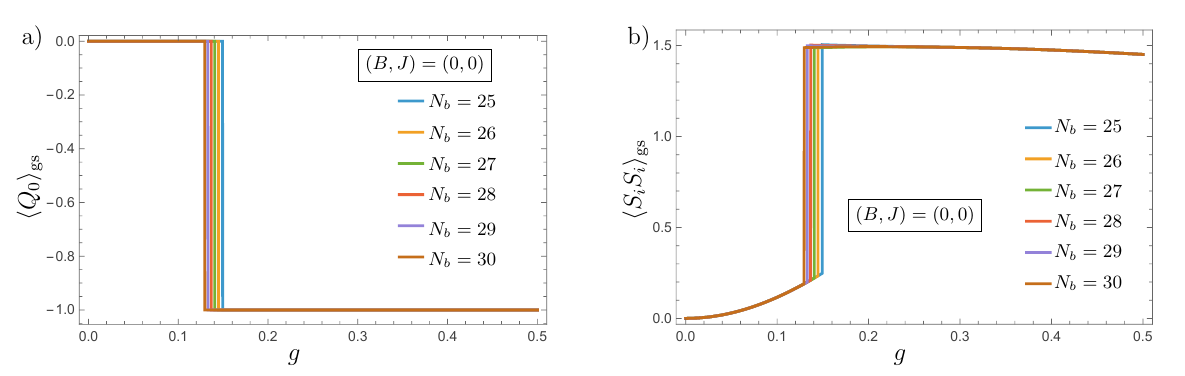}
\caption{The ground state expectation values  a) $\langle Q_0\rangle_{\textrm{gs}}$ and b) the quark spin-squared $\langle S^2\rangle_{\textrm{gs}}$ vs $g$ in the $(B,J)=(0,0)$ sector.  }\label{Fig_4}
\end{center}
\end{figure}
Consequently, in the vicinity of the critical point, the limiting behavior of $\Phi(g)$ on the delocalized side (i.e. for positive $(g-g_\ast^{00})$) is given by
\begin{eqnarray}
\lim_{N_b \to \infty} \Phi(g) \simeq  \frac{1}{\alpha (g-g_\ast^{00})^2}, \quad\quad \alpha= (0.285\pm 0.001) \quad\text{for }\,\, g\geq g_\ast^{00}. 
\end{eqnarray}
Thus $\Phi$ diverges at $g_\ast^{00}$, which provides  a striking characterization of the critical behavior.    
This divergence further substantiates the localization--delocalization transition inferred from $\mathcal{I}^{00}$.  Moreover, it plays a crucial role in revealing the nature of the ground state at the critical point, which we will discuss in a later section.

Alongside the change in the localization properties of the ground state, this QPT in the $(B,J)=(0,0)$ sector is accompanied by a change in its quark content. This change is captured by the ground state expectation value of axial charge $\langle Q_0\rangle_{\textrm{gs}}$, which exhibits a discontinuity at $\widetilde{g}^{00}(N_b)$ for each $N_b$ (Fig.\ref{Fig_4}a). Thus, in the converged limit, $\displaystyle{\lim_{{g\to g_\ast^{00}-}}}\langle Q_0\rangle_{\textrm{gs}} \neq \displaystyle{\lim_{g\to g_\ast^{00}+}}\langle Q_0\rangle_{\textrm{gs}}$ is an additional feature of the QPT.

 It is then natural to ask whether the fermionic component of the ground state undergoes a corresponding  corresponding restructuring of its spin composition at the QPT. 
 To investigate this, we examine the expectation value $\langle S_i S_i \rangle_{\textrm{gs}}$ as a function of $g$. 
 In general, an energy eigenstate can be decomposed into quark and glue components transforming in different irreps of $SO(3)_{rot}$: 
 \begin{eqnarray}
 |\Psi\rangle = \sum_{s,\ell} c_{s\ell} \sum_{s_3, \ell_3} C_{s, s_3; \ell, \ell_3}^{J, J_3} |s, s_3\rangle \otimes |\ell, \ell_3\rangle, \quad\quad \sum_{s,\ell}|c_{s\ell}|^2=1, \quad\quad c_{s\ell}\in \mathbb{C}. \label{gs_form}
 \end{eqnarray}
 Here, $s$ and $\ell$ denote the quark and glue spins, respectively, while $J$ is the total spin of the state. The $C_{s, s_3; \ell, \ell_3}^{J, J_3}$'s are the corresponding Clebsch-Gordan coefficients. We have suppressed the color indices for brevity. For the ground state of $(B,J)=(0,0)$ sector, the ground state $ |\Psi_{\textrm{gs}}\rangle$ has $J=0=J_3$ and in this case, the allowed combinations are $(s, \ell)$ are $\{(s, \ell)\}=\{(0,0), (1,1), (2,2)\}$. Consequently, 
 \begin{eqnarray}
 \langle S_i S_i \rangle_{\textrm{gs}}= 2|c_{11}|^2+6 |c_{22}|^2. 
 \end{eqnarray}

 At very small $g$, the ground state is predominantly composed of a spin-0 quark component coupled to a spin-0 glue configuration, and $|c_{00}|^2$ close to unity. The contributions from the higher-spin components appears as  small perturbative corrections, so that $|c_{11}|^2$ and $|c_{22}|^2$ remain small. Our numerical results indeed confirm this behavior throughout the localized phase, $g<g_\ast^{00}$, as shown in Fig.~\ref{Fig_4}b. The perturbative description breaks down at the transition, and the higher-spin quark components acquire substantially larger weights  (i.e. large $|c_{11}|^2$ and $|c_{22}|^2$) in the ground state for  $g>g_\ast^{00}$. Accordingly, $\langle S_i S_i \rangle_{\textrm{gs}}$ is enhanced in the delocalized phase and exhibits a discontinuity at the transition.

One may be tempted to think that the discontinuities in the expectation values $\langle Q_0\rangle_{\textrm{gs}}$ and $\langle S_i S_i \rangle_{\textrm{gs}}$ are necessary features of the QPT and the localization-delocalization transition always induces a restructuring of the fermionic component of the ground state as well. However, it is important to emphasize that it is not the case. As we show below,  the localization-delocalization transition occurs in the other $(B,J)$ sectors without an accompanying discontinuity in  $Q_0$ and/or $ S_i S_i $.

We note that computing the localization measure $\mathcal{I}^{BJ}$ is numerically demanding, restricting our simulations to a maximum bosonic cutoff of $N_b=23$. In contrast, for observables
such as the ground-state energy $E_0$ and $\Phi$, we can reach $N_b=30$. Accordingly, we determine the critical couplings $g_\ast^{B,J}$ from the sharp transitions in $\mathcal{I}^{BJ}$ for
$N_b\leq23$, and use the peaks in $\Phi$ to estimate them for $24\leq N_b\leq30$.

\begin{figure}[t]
\begin{center}
\includegraphics[width=16cm]{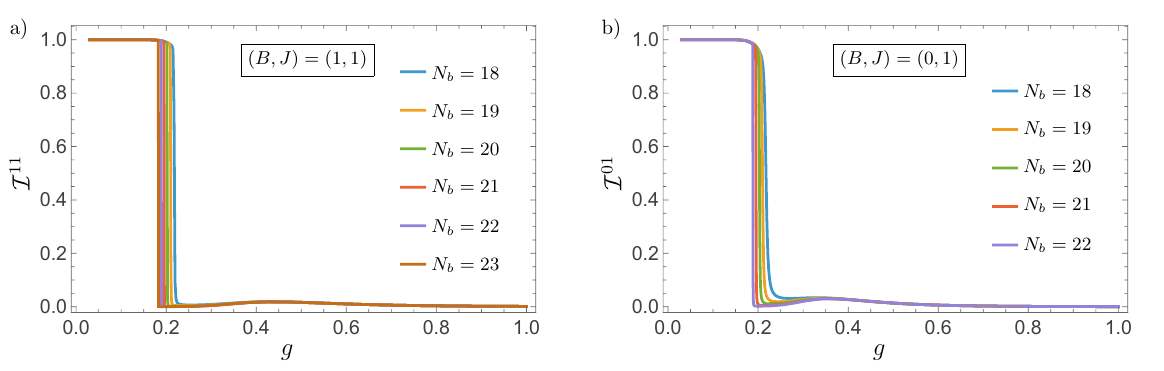}
\caption{The The localization measure a) $\mathcal{I}^{11}$ in $(B,J)=(1,1)$ and b) $\mathcal{I}^{01}$ in $(B,J)=(1,1)$ vs $g$ for different $N_b$.  }\label{Fig_5}
\end{center}
\end{figure}

\begin{figure}[h]
\begin{center}
\includegraphics[width=16cm]{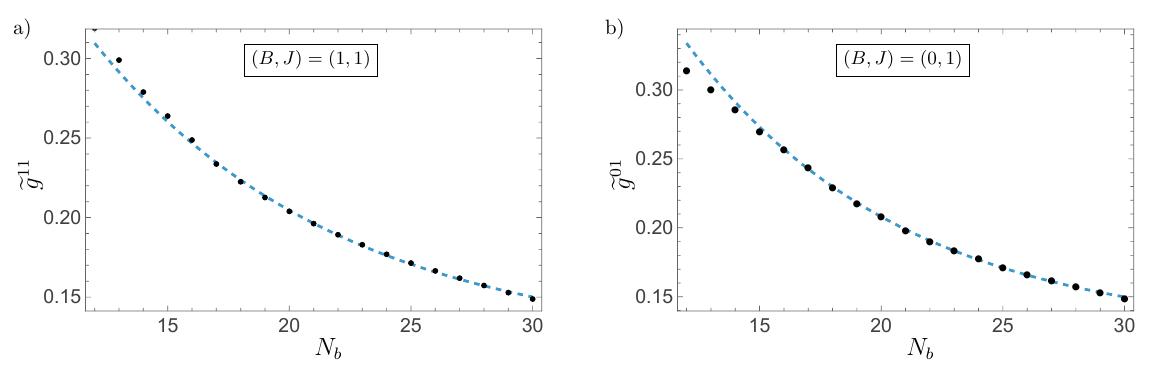}
\caption{a) $\widetilde{g}^{11}$ and b) $\widetilde{g}^{01}$ vs $g$ for different $N_b$. The dashed lines represent the fits in (\ref{fit_g_1101}). }\label{Fig_6}
\end{center}
\end{figure}

\subsection{$(B, J)=(1,1)$ and $(B, J)=(0,1)$ sectors}
We now turn to the $(B, J)=(1,1)$  sector. The ground state in this sector is a ninefold degenerate: it is spin-triplet with $J_3=-1,0,1$ and $SU(2)_B$-triplet  with $B_3=-1,0,1$. For the ground state $|\Psi_{\mathrm{gs}}^{J_3, B_3}\rangle$, we can define the localization measure in the $(B, J)=(1,1)$ sector as
\begin{eqnarray}
(B, J)=(1,1) \textrm{ sector:} \quad\quad \mathcal{I}^{11} \equiv \int_\Gamma\rho_{_\mathcal{G}}^{J_3}(M) \,d\mu_M, \quad\quad \rho^{J_3}_{_\mathcal{G}}(M)\equiv \langle\Psi_{\mathrm{gs}}^{J_3, B_3}|\Psi_{\mathrm{gs}}^{J_3, B_3}\rangle_{_F}  
\end{eqnarray}
This localization measure $\mathcal{I}^{11}$ is independent of both $J_3$ and $B_3$ (see  Appendix~\ref{J3B3_ind_I11} for details).  
 
Likewise in the $(B,J)=(0,1)$ sector, the ground state is a $SU(2)_B$ singlet and  a spin-1 triplet. It is therefore threefold degenerate, with states  labelled by $J_3=-1,0,1$.  
For the ground state $|\Psi_{\mathrm{gs}}^{J_3}\rangle$, we define define 
\begin{eqnarray}
(B, J)=(0,1) \textrm{ sector:} \quad\quad \mathcal{I}^{01} \equiv \int_\Gamma\rho_{_\mathcal{G}}^{J_3}(M) \,d\mu_M, \quad\quad \rho^{J_3}_{_\mathcal{G}}(M)\equiv \langle\Psi_{\mathrm{gs}}^{J_3}|\Psi_{\mathrm{gs}}^{J_3}\rangle_{_F}.  
\end{eqnarray}
\begin{figure}[t]
\begin{center}
\includegraphics[width=16cm]{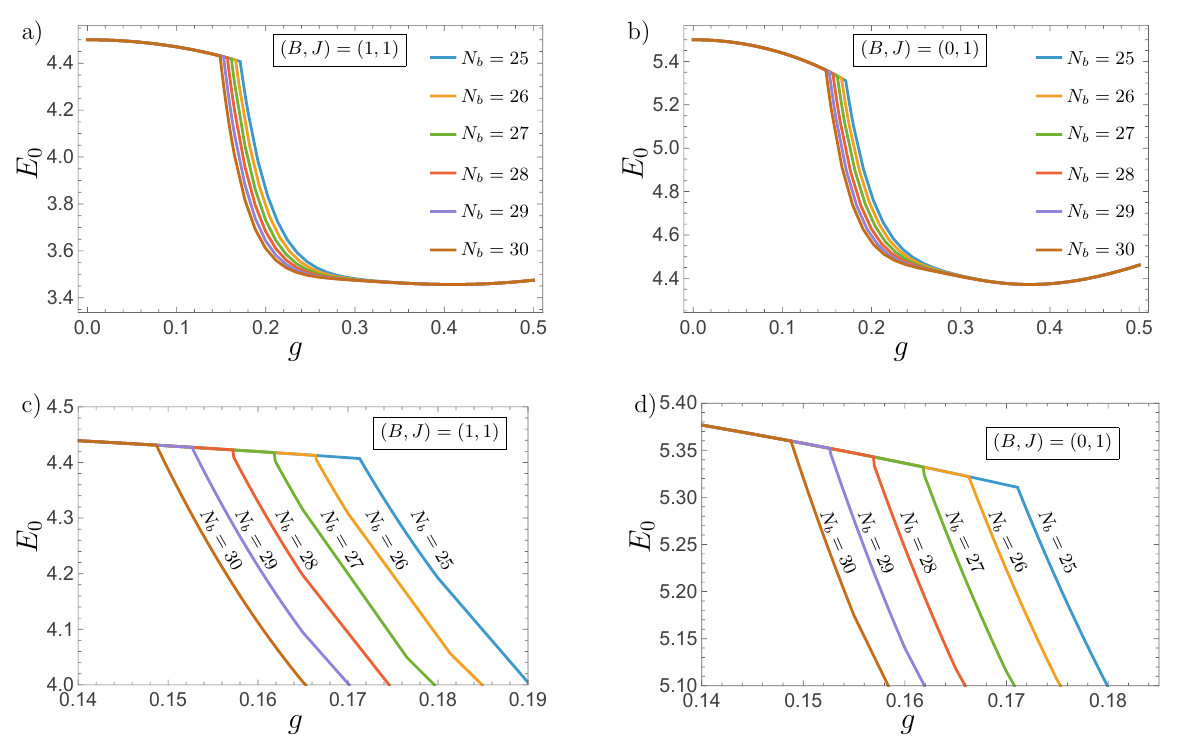}
\caption{The ground state energy  vs $g$ in a) $(B,J)=(1,1)$ and b) $(B,J)=(0,1)$ sectors. c) and d) correspond to zooming in the region of the kink.   }\label{Fig_7}
\end{center}
\end{figure}
\begin{figure}[h]
\begin{center}
\includegraphics[width=16cm]{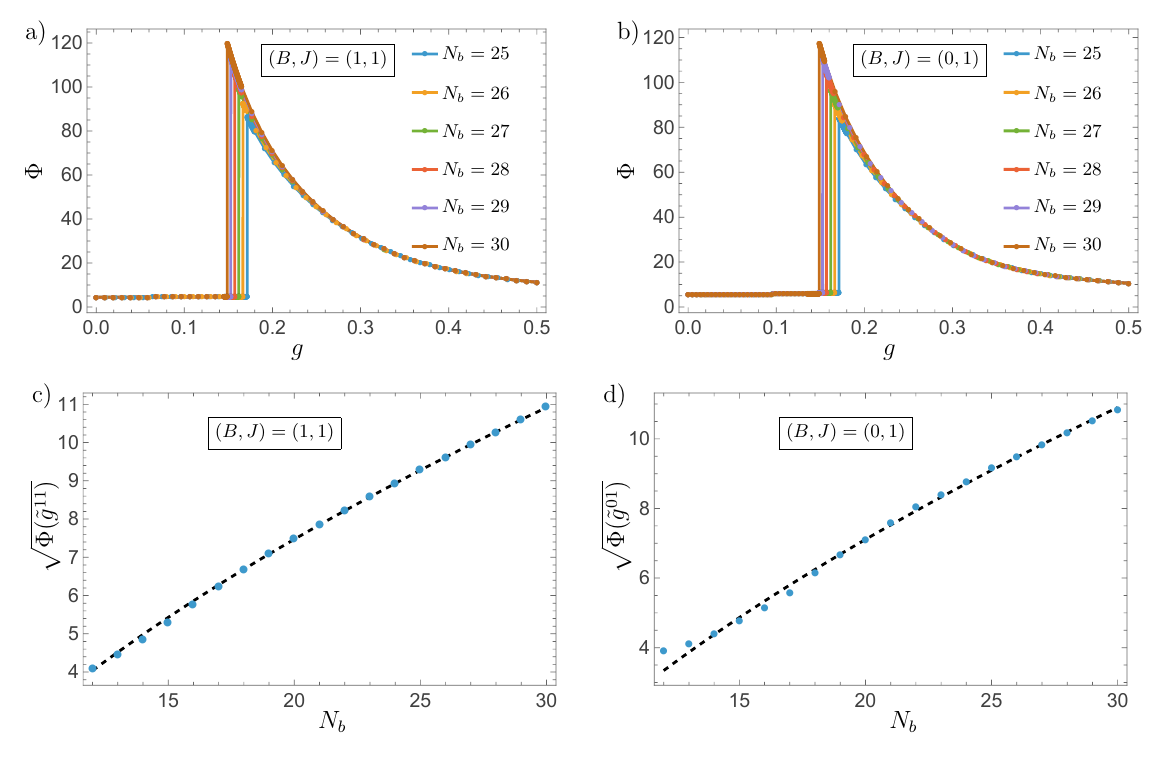}
\caption{$\Phi$ vs $g$ for various $N_b$ in the a) $(B,J)=(1,1)$ and b) $(B,J)=(0,1)$ sectors. The peak value $\sqrt{\Phi(\widetilde{g}^{BJ})}$ vs $N_b$ are shown in a) $for (B,J)=(1,1)$ and b) for $(B,J)=(0,1)$ sectors. The black dashed lines correspond the fits in (\ref{fit_phi_B1J1_B0J1}). }\label{Fig_8}
\end{center}
\end{figure}
For the ground state of the $(B,J)=(1,1)$ and $(B,J)=(0,1)$ sectors, the plots of $\mathcal{I}^{11}$ and $\mathcal{I}^{01}$ in the weak coupling regime for different values of the cut-off  $N_b$ are shown in Fig.~\ref{Fig_5}a and Fig.~\ref{Fig_5}b.  Again, we find that there exist a coupling $\widetilde{g}^{BJ}(N_b)$ for each of these two sectors such that 
\begin{eqnarray}
\mathcal{I}^{BJ}(g,N_b)  \left\{ \begin{array}{ll}
\simeq  1, \quad & \text{for } g< \widetilde{g}^{BJ}(N_b), \\ [0.75 ex]
\ll 1,  \quad & \text{for } g> \widetilde{g}^{BJ}(N_b). 
\end{array}\right.
\end{eqnarray}
Interestingly, we observe that 
\begin{eqnarray}
\lim_{\epsilon \to 0^+} \mathcal{I}^{BJ}( \widetilde{g}^{BJ}+ \epsilon ,N_b)\simeq 0,
\end{eqnarray}
for all $N_b$, while it  may, in general, have a  very small   non-zero value for $g>\widetilde{g}^{BJ}(N_b)$. 

For both these sectors, $\widetilde{g}^{BJ}$ as a function of $N_b$ is shown in (Fig.~\ref{Fig_6}a-b).  These can be fitted well by 
\begin{eqnarray}
\widetilde{g}^{BJ}(N_b) \simeq g_\ast^{BJ}  - c e^{- k N_b} \label{fit_g_1101}
\end{eqnarray}
with the values of $g_\ast^{BJ}$, $c$ and $k$ in Table~\ref{Table_1}. 
Thus,  in the $N_b \to \infty$ limit, $\widetilde{g}^{BJ}(N_b)$ converges to $g_\ast^{BJ}$ and  for both $(B,J)=(1,1)$ and $(B,J)=(0,1)$ sectors
\begin{eqnarray}
\lim_{N_b \to \infty} \mathcal{I}^{BJ}(g, N_b) \left\{ \begin{array}{ll}
\simeq  1, \quad & \text{for } g<g_\ast^{BJ}, \\[0.75 ex]
\ll 1,  \quad & \text{for } g>g_\ast^{BJ}. 
\end{array}\right.
\end{eqnarray}
Thus, in the $(B,J)=(1,1)$ and $(0,1)$ sectors too, there are QPTs at $g_\ast^{BJ}$ between a localized phase and a delocalized  phase.

As in the
$(B,J)=(0,0)$ sector, the energy drops sharply immediately
above the kink, with its slope becoming progressively
steeper as $N_b$ increases. This behavior suggests that,
in the converged limit, the ground-state energies become
non-analytic at their respective critical couplings
$g_\ast^{BJ}$.

In both these sectors, the ground state energy $E_0$ exhibits a kink at $\widetilde{g}^{BJ}(N_b)$ for every value of $N_b$ (Fig.~\ref{Fig_7}a-b). As in the $(B,J)=(0,0)$ sector, here too, the ground state energy drops sharply for $g_{\ast}^{B,J}+\epsilon$ (with $\epsilon \to 0^+$),  with its slope becoming progressively steeper as $N_b$ increases. This behavior suggests that,  in the converged limit, the ground state energies of the $(B,J)=(1,1)$ and $(0,1)$ sectors become non-analytic at $g_\ast^{BJ}$.

The observable $\Phi(g)$  also  has  a sharp peak at $\widetilde{g}^{BJ}(N_b)$ in both sectors, as shown in Fig.~\ref{Fig_8}a-b. The peak heights  $\sqrt{\Phi(\widetilde{g}^{BJ})}$ progressively increase with $N_b$, and their dependence on $N_b$ can be well described by
\begin{eqnarray}
\sqrt{\Phi(\widetilde{g}^{BJ})} \simeq A+ B N_b^{\gamma}.  \label{fit_phi_B1J1_B0J1}
\end{eqnarray}
with parameters $A$, $B$ and $\gamma$ given in Table~\ref{Table_2}. This implies that as $N_b \to \infty$, the peak heights, and hence $\Phi(g_\ast^{BJ})$ diverges as a power law for both $(B,J)=(1,1)$ and $(B,J)=(0,1)$ sectors.

\begin{figure}
\begin{center}
\includegraphics[width=18cm]{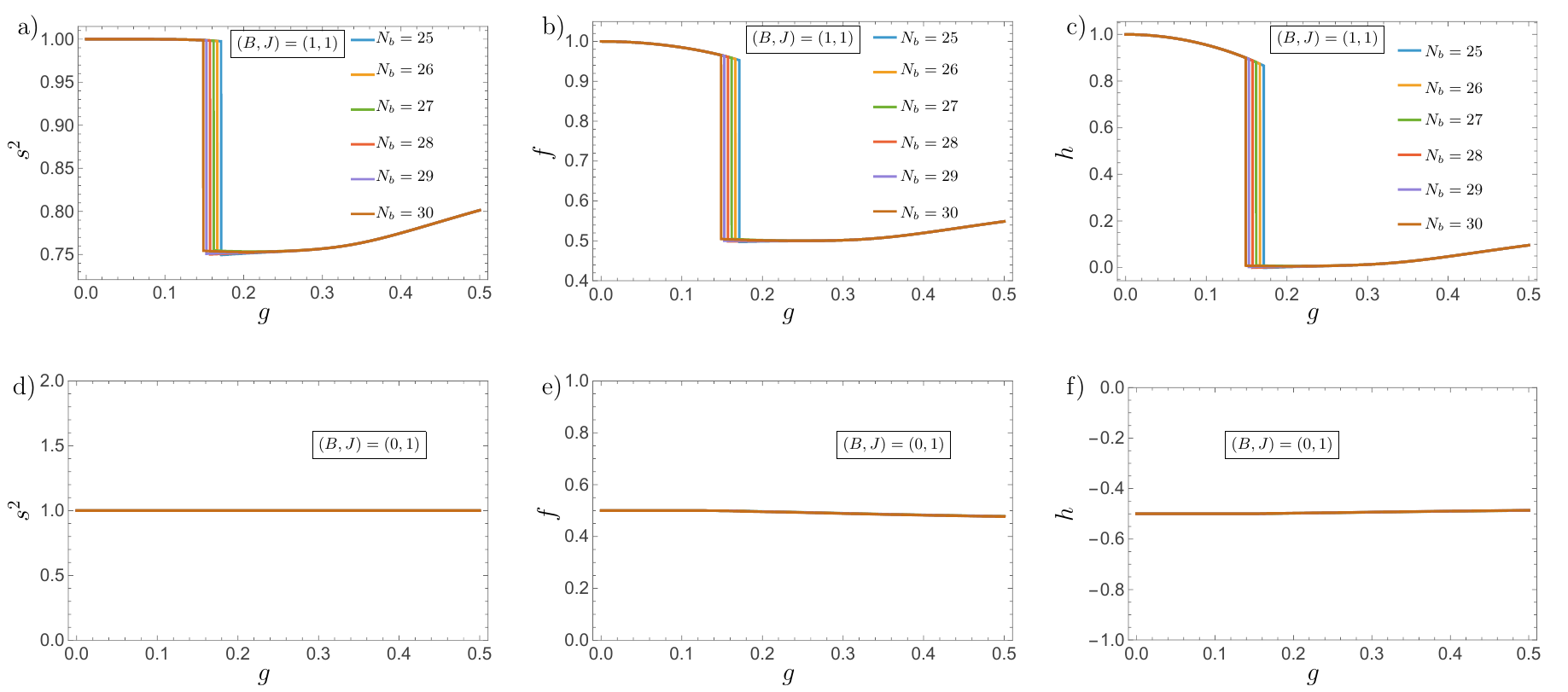}
\caption{$s^2$, $f$ and $h$ as functions of $g$.  a-c) For $(B,J)=(1,1)$ sector.  d-f) For $(B,J)=(0,1)$ sector.}\label{Fig_9}
\end{center}
\end{figure}

We now turn to the quark content of the ground states and examine whether they undergo a change at these QPTs.  In contrast to the $(B,J)=(0,0)$ sector, we find that the axial charge  $\langle Q_0 \rangle_{\textrm{gs}} $ remains a constant throughout the weak coupling regime in both sectors: 

\begin{eqnarray}
\langle Q_0 \rangle_{\textrm{gs}}= \left\{
\begin{array}{ll}
0& \quad \text{ for }    (B,J)=(1,1)\\ [0.75 ex]
-1 & \quad \text{ for }  (B,J)=(0,1). 
\end{array}
\right. 
\end{eqnarray}
Thus, unlike the $ (B,J)=(0,0)$ sector, neither of these two sectors exhibits a discontinuity in the axial charge across the localization-delocalization transition at $g_\ast^{BJ}$.

The division of the spin of the ground state between the quark and glue in these sectors is particularly interesting.  Since the ground states in these sectors have total spin $J=1$, the expectation value $\langle S_i S_i \rangle_{\textrm{gs}}$ alone does not provide a  complete characterization of the quark-spin structure. We therefore also consider the expectation values of quark spin $\langle S_i \rangle_{\textrm{gs}}$ and the quadrupole spin tensor $\langle Q_{ij}^{(S)} \rangle_{\textrm{gs}}=\langle(S_i S_j+ S_jS_i -\frac{2}{3} \delta_{ij} S_kS_k) \rangle_{\textrm{gs}}$. By the Wigner-Eckart theorem, within a fixed fixed $J=1$ multiplet, these expectation values are related to the corresponding total-spin observables as
\begin{eqnarray}
\langle S_i S_i \rangle_{\textrm{gs}}= s^2 \langle J_i J_i \rangle_{\textrm{gs}}= 2 s^2, \quad\quad \langle S_i \rangle_{\textrm{gs}}= f \langle J_i \rangle_{\textrm{gs}}, \quad\quad \langle Q_{ij}^{(S)} \rangle_{\textrm{gs}}= h \langle  Q_{ij}^{(J)} \rangle_{\textrm{gs}}, \quad s^2, f, h \in \mathbb{R}, \label{spin_relations}
\end{eqnarray}
where $\langle Q_{ij}^{(J)} \rangle_{\textrm{gs}}=\langle(J_i J_j+ J_jJ_i -\frac{2}{3} \delta_{ij} J_kJ_k) \rangle_{\textrm{gs}}$.  Here, $s^2$, $f$, and $h$ are real coefficients characterizing the quark-spin content of the state and these relations holds true for any  $J=1$ state which can be an arbitrary linear combination of components with different  $J_3$.

As in (\ref{gs_form}), the ground states can be decomposed into quark and gluonic spin components with the allowed $(s,\ell)$ combinations: 
\begin{eqnarray}
\{(s,\ell)\}= \left\{
\begin{array}{ll}
\{ (0,1), (1,0), (1,1),(1,2) \}, & \quad \text{ for }    (B,J)=(1,1)\\ [0.75 ex]
\{ (0,1), (1,1), (1,2),(2,2) \}, & \quad \text{ for }    (B,J)=(0,1).
\end{array}
\right.
\end{eqnarray}
In terms of the corresponding amplitudes $c_{s\ell}$, the coefficients $s^2$, $f$, and $h$ in (\ref{spin_relations}) are given by
\begin{eqnarray}
&& \text{ for }    (B,J)=(1,1): \quad \left\{
\begin{array}{ll} 
s^2=|c_{10}|^2+|c_{11}|^2+|c_{12}|^2,\\ [0.75 ex]
 f=|c_{10}|^2+\frac{1}{2} |c_{11}|^2-\frac{1}{2} |c_{12}|^2, \\[0.75 ex]
  h=|c_{10}|^2-\frac{1}{2}|c_{11}|^2+\frac{1}{10}|c_{12}|^2
\end{array}\right.  \\
&& \text{ for }    (B,J)=(0,1): \quad \left\{
\begin{array}{ll} 
s^2=|c_{11}|^2+|c_{12}|^2+ 3 |c_{2,2}|^2,\\ [0.75 ex]
 f=\frac{1}{2} |c_{11}|^2-\frac{1}{2} |c_{12}|^2+ \frac{1}{2} |c_{22}|^2, \\[0.75 ex]
  h=-\frac{1}{2}|c_{11}|^2+\frac{1}{10}|c_{12}|^2- \frac{7}{5} |c_{22}|^2.
\end{array}\right.  
\end{eqnarray}
The normalization of the state requires $0\leq |c_{s\ell}|\leq 1$ and $\sum_{s,\ell}|c_{s\ell}|^2=1$, which imposes the following kinematic bounds:
\begin{eqnarray}
&& \text{ for }    (B,J)=(1,1): \quad 0\leq s^2 \leq 1, \quad\quad -\frac{1}{2} \leq f \leq 1, \quad\quad -\frac{1}{2} \leq h \leq 1, \\
&& \text{ for }    (B,J)=(0,1): \quad 0\leq s^2 \leq 3, \quad\quad -\frac{1}{2} \leq f \leq 1, \quad\quad -\frac{7}{5} \leq h \leq \frac{1}{10}. 
\end{eqnarray}
 The actual values of $s^2, f$ and $h$ are dynamically determined, and  their dependance on $g$ for the $(B,J)=(1,1)$ and $(B,J)=(0,1)$ sectors are shown in Fig.~\ref{Fig_9}.

For very small $g$ in the $(B,J)=(1,1)$ sector, the ground state is primarily composed of spin-1 quark state coupled to spin-0 glue state,  with small perturbative corrections from the other $(s, \ell)$ states. In other words, the coefficient $|c_{10}|$ is close to unity, which yields $s^2$, $f$ and $h$ close to one for $g<\widetilde{g}^{11}(N_b)$. This behavior changes sharply at $\widetilde{g}^{11}(N_b)$ and the other $(s,\ell)$ components acquire substantial weights in the ground state  for  $g>\widetilde{g}^{11}(N_b)$. This redistribution of the spin among the quark and the glue  leads to a sharp decrease in $s^2$, $f$, and $h$. Thus, in the converged limit, the quark spin structure undergoes a discontinuous change at $g_\ast^{11}$.

 The behavior of the quark spin across the QPT in $(B,J)=(0,1)$ sector is markedly different. Here,  $s^2, f$ and $h$ all remain continuous at the  $g_\ast^{01}$ (Fig.~\ref{Fig_9}d-f).  In fact, we find that  $s^2 \simeq 1$, $f\simeq \frac{1}{2}$ and $h\simeq -\frac{1}{2}$  throughout the weak coupling regime. These values suggest that the ground state of this sector is almost entirely composed of $(s, \ell) =(1,1)$ component,  both in the localized and delocalized phases .

These results establish that  localization-delocalization transitions occur in the  $(B,J)=(1,1)$ and $(B,J)=(0,1)$ sectors as well. However, these transitions do not necessarily entail a restructuring of the fermionic component of the ground state. Such a restructuring happens in the $(B,J)=(1,1)$ sector but is absent in the $(B,J)=(0,1)$ sector.

\begin{figure}[t]
\begin{center}
\includegraphics[width=16cm]{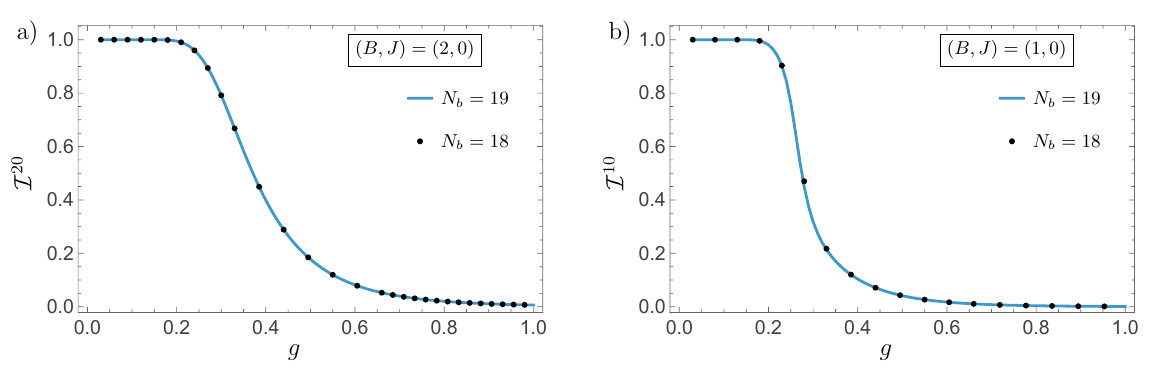}
\caption{a) $\mathcal{I}^{2,0}$ and b) $\mathcal{I}^{1,0}$ vs $g$. The black dots correspond to $N_b=18$ and the solid lines correspond to $N_b=19$.   }\label{Fig_10}
\end{center}
\end{figure}

\subsection{$(B, J)=(1,0)$ and $(B, J)=(2,0)$ sectors}

In the  $(B, J)=(2,0)$ sectors, all states are of the form spin-0 quark state $\otimes$ spin-0 glue state. On the other hand, the states in the  $(B, J)=(1,0)$ sectors have total spin $J=0$ and are of the form spin-1 quark state $\otimes$ spin-1 glue state.  Thus, the ground states in these sectors have the following spin compositions: 
\begin{eqnarray}
\{(s,\ell)\}= \left\{
\begin{array}{ll}
\{ (0,0)  \}, & \quad \text{ for }    (B,J)=(2,0)\\ [0.75 ex]
\{  (1,1)\}, & \quad \text{ for }    (B,J)=(1,0).
\end{array}
\right.
\end{eqnarray}
As a result, the quark spin component in these sectors are fixed: 
\begin{eqnarray}
\langle S_i S_i \rangle_{\textrm{gs}}= \left\{
\begin{array}{ll}
0 & \quad \text{ for }    (B,J)=(2,0)\\ [0.75 ex]
2 & \quad \text{ for }    (B,J)=(1,0). 
\end{array}
\right.
\end{eqnarray}
The quark spin of the ground states in these sectors are therefore independent of the coupling $g$ and remain continuous throughout the weak coupling regime.

\begin{figure}[h]
\begin{center}
\includegraphics[width=16cm]{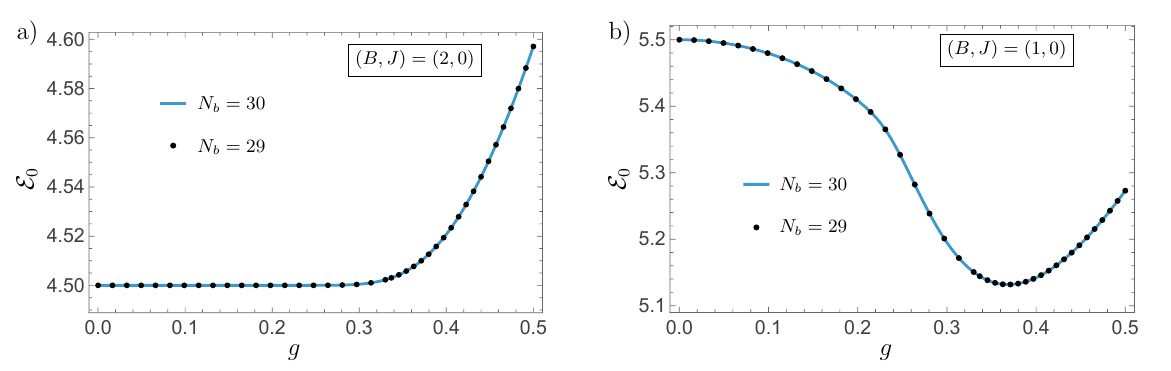}
\caption{The lightest energy of a) $(B,J)=(2,0)$ and b)  $(B,J)=(2,0)$ sectors  in the weak coupling regime.  The black dots correspond to $N_b=29$ and the solid lines correspond to $N_b=30$.   }\label{Fig_11}
\end{center}
\end{figure}

Furthermore, the states in both these sectors always have four fermions (quarks or anti-quarks depending on $B_3$). The axial charge of the ground states of these sectors is thus always $\langle Q_0\rangle_{\textrm{gs}}=0$.

\begin{figure}[h]
\begin{center}
\includegraphics[width=16cm]{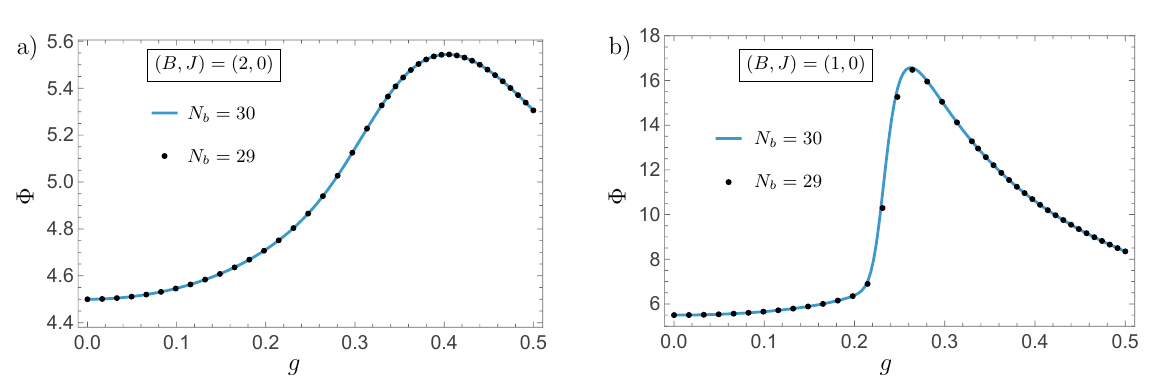}
\caption{The ground state expectation value $\Phi$ for a) $(B,J)=(2,0)$ and b)  $(B,J)=(2,0)$  sectors in the weak coupling regime.  The black dots correspond to $N_b=29$ and the solid lines correspond to $N_b=30$.   }\label{Fig_12}
\end{center}
\end{figure}

We can compute the  localization measure $\mathcal{I}^{20}$ and $\mathcal{I}^{10}$ in these sectors as well. We find that both these quantities vary continuously throughout $g \in[0,1]$ (Fig.~\ref{Fig_10}a-b). The ground state energies likewise remain analytic  in the entire weak coupling regime (Fig.~\ref{Fig_11}a-b), while the ground state expectation values $\Phi(g)$ also remain finite (Fig.~\ref{Fig_12}a-b). These observations indicate that neither of these two sectors encounter a localization-delocalization transition in the weak coupling regime.

One might ask, what is fundamentally different in these two sectors from the previous ones? The answer lies in the quark-glue interaction. For both the $(B, J)=(1,0)$ and $(B, J)=(2,0)$ sectors, the matrix elements of the quark-glue interaction term vanishes identically. As the quark and the glue do not interact, the states in these sectors are pure glueball-like, accompanied by four fermions that adjust the quantum numbers of the physical state. More specifically, the states in $(B, J)=(2,0)$ sector are spin-0 colorless glueballs accompanied by a spin-0 colorless four-fermion state. The difference between the numbers of quarks and anti-quarks gives the $B_3$, which in this case can take integer values  $B_3=-2,-1,0,1,2$. In contrast, the states in $(B, J)=(1,0)$ sector contain spin-1 color-triplet glueballs which combine with spin-1 color-triplet 4-fermion states to form an overall entangled color-singlet with spin-0.

These results highlight the importance of quark-glue interaction as a necessary ingredient in the  localization-delocalization transition. The transition occurs only in sectors with non-trivial interaction. 
It is worth noting, however, that the $(B,J)=(1,0)$ sector exhibit a relatively sharp crossover compared to the  $(B,J)=(2,0)$ sector. This is reflected in the rapid, although continuous, variation of $\mathcal{I}^{10}$ and $\Phi$ over the coupling range $0.2 \lesssim g \lesssim 0.4$. 

\begin{figure}[h]
\begin{center}
\includegraphics[width=16cm]{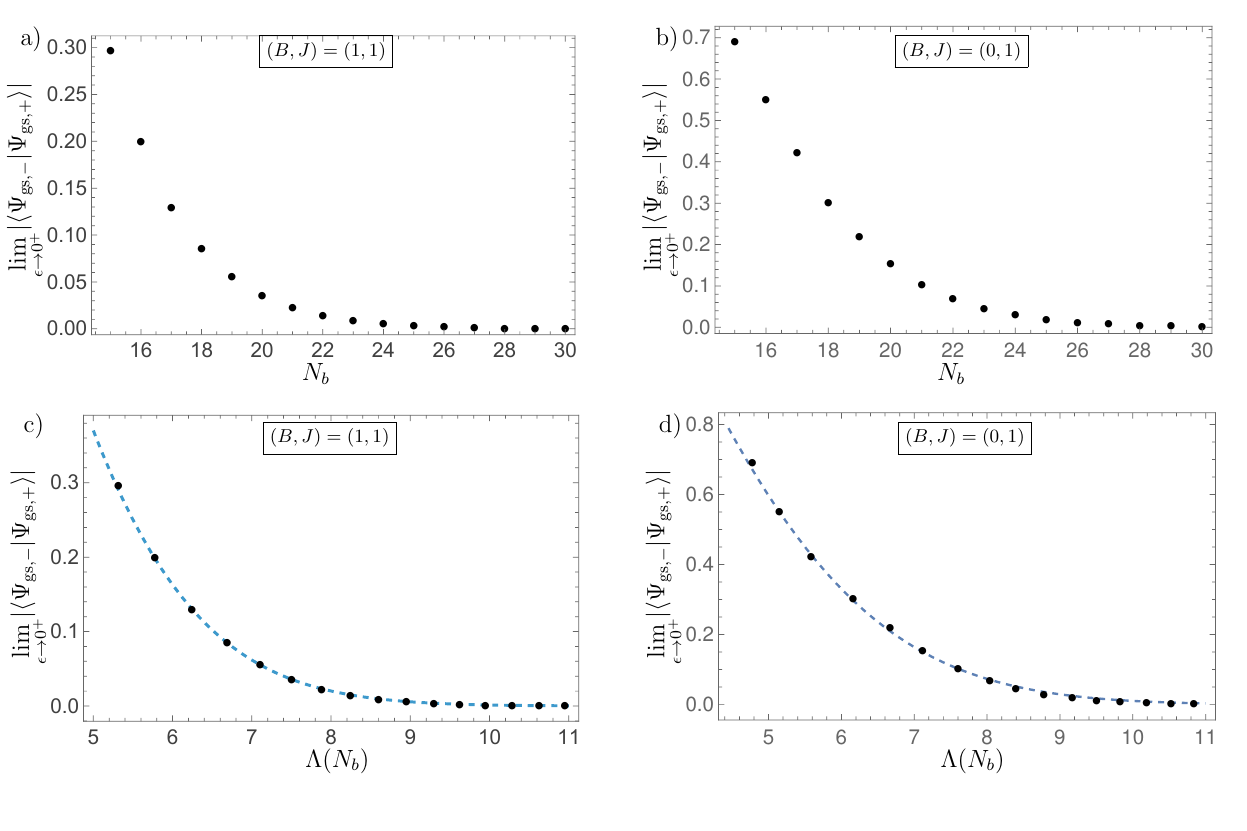}
\caption{The plot of $\displaystyle{\lim_{\epsilon \to 0^+} }|\langle \Psi_{\textrm{gs},-}|\Psi_{\textrm{gs},+}\rangle|$ vs $N_b$ for a) $(B,J)=(1,1)$ and b) $(B,J)=(0,1)$ sectors. The variation of $\displaystyle{\lim_{\epsilon \to 0^+} }|\langle \Psi_{\textrm{gs},-}|\Psi_{\textrm{gs},+}\rangle|$ vs $\Lambda(N_b)$   for c) $(B,J)=(1,1)$ and d) $(B,J)=(0,1)$ sectors. }\label{Fig_13}
\end{center}
\end{figure}

\subsection{Critical phenomena}\label{critical_phono}

In this section, we investigate the nature of the ground state wavefunctions in the $(B,J)=(0,0), (1,1)$ and $(0,1)$ sectors in the vicinity of their respective critical couplings $g_\ast^{BJ}$. With $0\leq \epsilon\ll 1$, we define the ground state wavefunctions on either side of the critical point as 
\begin{eqnarray}
|\Psi_{\textrm{gs},\pm}\rangle \equiv |\Psi_{\textrm{gs}} (g_\ast^{BJ}\pm \epsilon) \rangle
\end{eqnarray} 
The analysis in the previous sections shows that, as $\epsilon\to 0^+$, the state $|\Psi_{\textrm{gs},-}\rangle$ becomes almost completely localized inside the region $\Gamma$, whereas  $|\Psi_{\textrm{gs},+}\rangle$ has negligible probability in $\Gamma$.

\textit{Superselection sectors:} To characterize the increasing extent of the delocalized wavefunction, we introduce the scale
\begin{eqnarray}
\Lambda^2(\epsilon) \equiv \frac{2}{R^2} \langle \Psi_{\textrm{gs},+}| \text{Tr}(A_i A_i)|\Psi_{\textrm{gs},+}\rangle. 
\end{eqnarray}
From the results of the previous sections, we find that in the $(B,J)=(0,0), (1,1)$ and $(0,1)$ sectors, 
\begin{eqnarray}
\lim_{\epsilon \to 0^+} \Lambda^2(\epsilon) = \infty, \quad\quad   \lim_{\epsilon \to 0^+}\frac{2}{R^2} \langle \Psi_{\textrm{gs},-}|\text{Tr}(A_i A_i) |\Psi_{\textrm{gs},-}\rangle \ll \lim_{\epsilon \to 0^+} \Lambda^2(\epsilon). 
\end{eqnarray}

Thus, while the characteristic extent of the localized wavefunction $|\Psi_{\textrm{gs},-}\rangle$ in the gauge configuration space remains finite, that of the delocalized wavefunction $|\Psi_{\textrm{gs},+}\rangle$ diverges in the critical limit.  Together with the vanishing probability inside $\Gamma$, this shows that the  wavefunction $|\Psi_{\textrm{gs},+}\rangle$  spreads out over gauge configuration space.

We next compute the overlap between  $|\Psi_{\textrm{gs},-}\rangle$ and $|\Psi_{\textrm{gs},+}\rangle$ in the $\epsilon \to 0$ limit. In the $(B,J)=(0,0)$ sector, this overlap vanishes identically because the two states have different axial charges. 
The situation is more interesting in the $(B,J)=(1,1)$ and $(0,1)$ sectors, where the states  $|\Psi_{\textrm{gs},-}\rangle$ and $|\Psi_{\textrm{gs},+}\rangle$ have the same axial charge and their overlap is not constrained to vanish by this selection rule.
 Nevertheless, we find that that in these sectors,  the overlap $\displaystyle{\lim_{\epsilon \to 0^+} }|\langle \Psi_{\textrm{gs},-}|\Psi_{\textrm{gs},+}\rangle|$ decreases monotonically with $N_b$ and converges to zero (Fig.\ref{Fig_13}a-b).

An even more interesting picture is obtained by plotting  $|\langle \Psi_{\textrm{gs},-}|\Psi_{\textrm{gs},+}\rangle|$  vs $\Lambda$, with both quantities evaluated in the   $\epsilon\to0^+$ limit at each value of $N_b$ (Fig.\ref{Fig_13}c-d). We find that the dependence of the overlap on $\Lambda$ is well described by

\begin{eqnarray}
\displaystyle{\lim_{\epsilon \to 0^+} }|\langle \Psi_{\textrm{gs},-}|\Psi_{\textrm{gs},+}\rangle| \simeq \kappa e^{- t^2\Lambda^2},  
\end{eqnarray}
where  $t$ and $k$ are 
\begin{eqnarray}
&& \text{for }  (B,J)=(1,1): \quad \kappa \simeq 2.39 \pm 0.04, \quad\quad t\simeq 0.273 \pm 0.001 \\
&& \text{for }  (B,J)=(0,1): \quad \kappa \simeq 2.29 \pm 0.05, \quad\quad t\simeq 0.232\pm0.003
\end{eqnarray}

\begin{figure}
\begin{center}
\includegraphics[width=7cm]{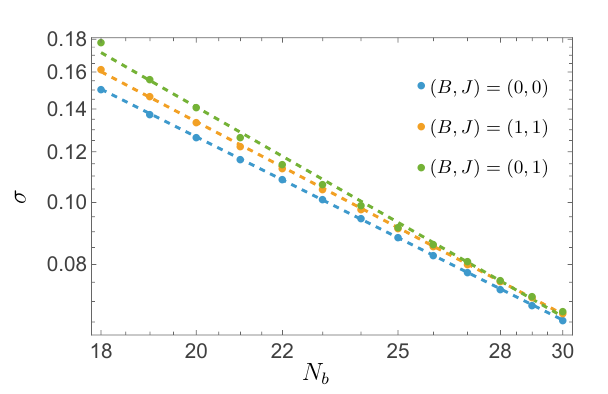}
\caption{Log-log plot of $\sigma$ vs $N_b$ in $(B,J)=(0,0)$, $(B,J)=(1,1)$, and  $(B,J)=(0,1)$ sectors. The dots correspond to numerical data and the dashed lines correspond to fits (\ref{fits_sigma}).}\label{Fig_15}
\end{center}
\end{figure}

The exponential suppression of this overlap has far-reaching implications. For any gauge invariant observable $\mathcal{O}$ which is a polynomial function of the canonical variables $A_i$ and $\Pi_i$, we expect  
\begin{eqnarray}
\lim_{\epsilon \to 0^+} |\langle \Psi_{\textrm{gs},-}|\mathcal{O}|\Psi_{\textrm{gs},+}\rangle| \sim (\Lambda^p + O(\Lambda^{p-1})) e^{- t^2\Lambda^2}
\end{eqnarray}
where $p$ is the degree of the polynomial.   As a result, in the converged limit (i.e. $N_b \to \infty$)
\begin{eqnarray}
\lim_{\epsilon \to 0^+} |\langle \Psi_{\textrm{gs},-}|\Psi_{\textrm{gs},+}\rangle|= 0, \quad\quad \lim_{\epsilon \to 0^+} |\langle \Psi_{\textrm{gs},-}|\mathcal{O}|\Psi_{\textrm{gs},+}\rangle|= 0. 
\end{eqnarray}
Thus, in the two limiting states become orthogonal to each other and decoupled with respect to the physically relevant gauge-invariant canonical observable algebra.

 Similar arguments can be invoked in the $(B,J)=(0,0)$ sector for the two limiting states  at $g_\ast^{00}$: there, the matrix elements of observables which preserve axial charge vanishes identically between the two limiting states and those which change the axial charge appropriately, vanish by exponential suppression.

These in turn,  provide evidence that, at the critical couplings $g_\ast^{BJ}$, the limiting ground states approached from the two sides of the transition belong to distinct superselection sectors with respect to the algebra of gauge-invariant observables.

\textit{Localization on configurations with equal singular values:} 
To investigate the structure of the ground state $| \Psi_{gs,+}\rangle$ in gauge configuration space, we introduce the dimensionless matrix $m_{ia} \equiv R\Lambda^{-1} M_{ia}$.    The real matrix $m_{ia}$ admits a singular value decomposition: $m=  \mathcal{R} D\mathcal{S}^T$ where $  \mathcal{R} \in O(3)_{rot}$, $\mathcal{S} \in  Ad(SU(2)) \simeq SO(3)_{col}$, and $D=\text{diag}(a_1,a_2,a_3)$.  Here, $a_i\in \mathbb{R}$ are the singular values of $m_{ia}$ and by construction,
\begin{eqnarray}
\langle \Psi_{gs,+ }| m_{ia} m_{ia}| \Psi_{gs,+ }\rangle= \langle \Psi_{gs,+ }| ({a}_1^2 +{a}_2^2+a_3^2) | \Psi_{gs,+ }\rangle=1. \label{Jensen_1}
\end{eqnarray}

To quantify the deviation from the surface of equal singular values, we define
\begin{eqnarray}
\sigma \equiv \frac{1}{2}\sum_{i,j=1}^3 \lim_{\epsilon \to 0^+} \langle \Psi_{gs,+ }| ({a}_i^2 -{a}_j^2)^2 | \Psi_{gs,+ }\rangle.
\end{eqnarray}
This quantity measures the mean-square differences between the squared singular values and can be evaluated directly in terms of the original configuration variables:
\begin{eqnarray}
\sigma =  \lim_{\epsilon \to 0^+} R^4 \Lambda^{-4} \Big[3\langle \Psi_{gs,+ }| M_{ia} M_{ja} M_{jb} M_{ib} | \Psi_{gs,+ }\rangle- \langle \Psi_{gs,+ }| M_{ia} M_{ia} M_{jb} M_{jb} | \Psi_{gs,+ }\rangle\Big].
\end{eqnarray}
Numerically, we find that $\sigma$ steadily decreases with $N_b$ in all three $(B,J)$ sectors (Fig.~\ref{Fig_15}) and  can be described as the power law decays: 
\begin{eqnarray}
\sigma \simeq \left\{ \begin{array}{ll}
(16.63 \pm 0.01) N_b^{-(1.628 \pm  0.002)}, \quad \text{for } (B,J)=(0,0),\\ [1 ex]
(22.43 \pm  0.03) N_b^{-(1.710 \pm  0.008)}, \quad \text{for } (B,J)=(1,1),\\ [1 ex]
(37.26 \pm  0.22) N_b^{-(1.86 \pm  0.03)}, \,\,\,\,\quad \text{for } (B,J)=(0,1). 
\end{array}
\right. \label{fits_sigma}
\end{eqnarray}
Thus as $N_b$ increases, $\sigma$  decreases and the limiting wavefunction $\displaystyle{\lim_{\epsilon \to 0^+}} | \Psi_{\textrm{gs},+} \rangle$ has progressively sharp concentration on the surface of equal singular values.
 
\begin{figure}[h]
\begin{center}
\includegraphics[width=18cm]{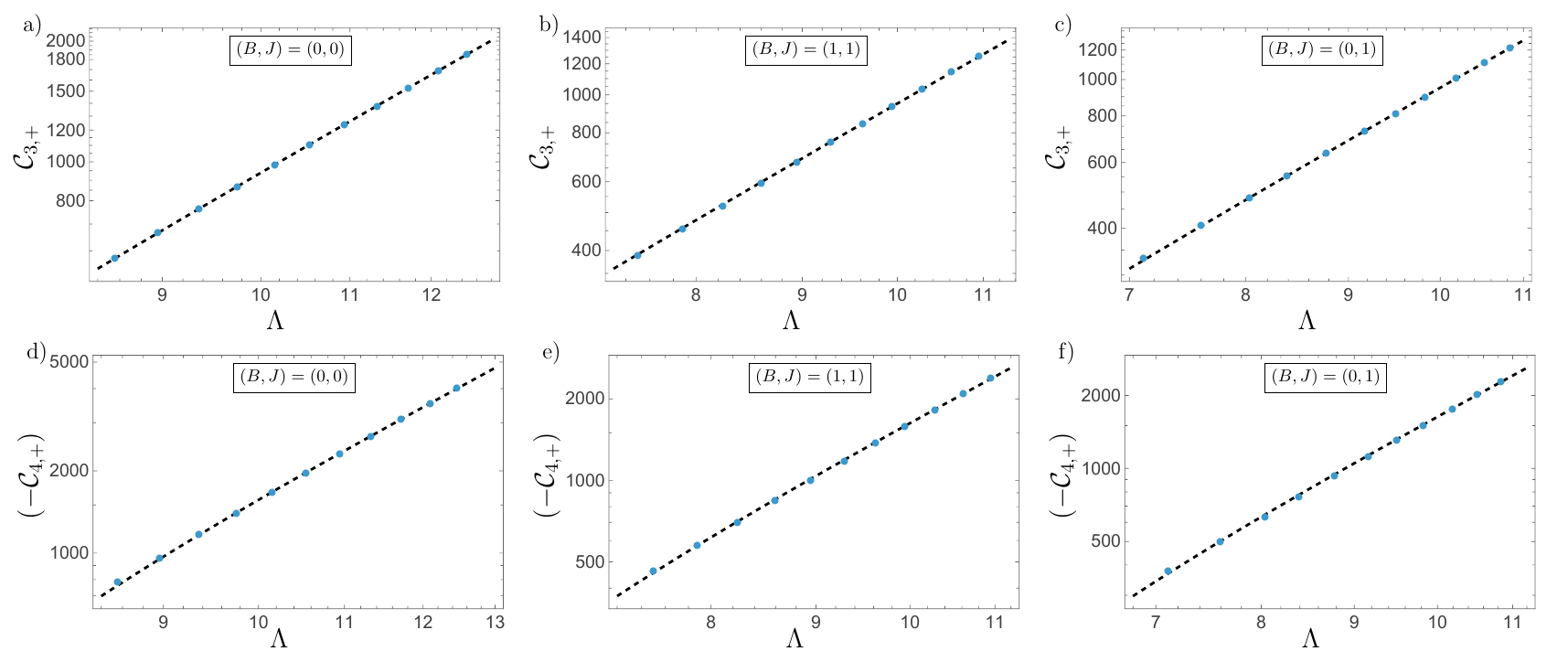}
\caption{Log-log plot of $\mathcal{C}_{3,+}$ vs $\Lambda$ in a) $(B,J)=(0,0)$, b)$(B,J)=(1,1)$, and c) $(B,J)=(0,1)$ sectors. d-f) The corresponding log-log plot of $(-\mathcal{C}_{4,+})$ vs $\Lambda$. The dots correspond to numerical data and the black dashed lines correspond to fits (\ref{fits_C3}).}\label{Fig_14}
\end{center}
\end{figure}

The normalization fixes the expectation value  $ \langle \Psi_{gs,+ }| ({a}_1^2 +{a}_2^2+a_3^2) | \Psi_{gs,+ }\rangle=1$, but does not, by itself, constrain its quantum fluctuations. To investigate these fluctuations, we consider the gauge-invariant quantities, which are analogues of the classical  cubic and quartic invariants: 
\begin{eqnarray}
&& \mathcal{C}_{3,+} \equiv  \frac{\sqrt{3}}{2R^3}  \epsilon_{ijk} \epsilon_{abc} \langle \Psi_{gs,+ }| M_{ia} M_{jb} M_{kc}| \Psi_{gs,+}\rangle, \\ 
&& \mathcal{C}_{4,+} \equiv \frac{9R^4}{16} \Big[2\langle \Psi_{gs,+ }| M_{ia} M_{ja} M_{jb} M_{ib} -\langle \Psi_{gs,+ }| M_{ia} M_{ia} M_{jb} M_{jb} | \Psi_{gs,+ }\rangle| \Psi_{gs,+ }\rangle\Big]. 
\end{eqnarray}

In the $\epsilon\to0^+$ limit, we compute $\mathcal{C}_{3,+}$ and $\mathcal{C}_{4,+}$ in the different $(B,J)$ sectors for various values of $N_b$ and plot them against the corresponding $\Lambda$ (Fig.~\ref{Fig_14}). The numerical results are well described by
\begin{equation}
\lim_{\epsilon\to0^+}\mathcal{C}_{3,+}\simeq A_3+B_3\Lambda^{\beta_3},
\qquad
\lim_{\epsilon\to0^+}\mathcal{C}_{4,+}\simeq -A_4-B_4\Lambda^{\beta_4},
\label{fits_C3}
\end{equation}
with the fit parameters in Table~\ref{Table_3}. In particular, we find $\beta_3 \simeq 3$ and$\beta_4 \simeq 4$, with $B_3 \simeq 1$ and $B_4 \simeq \frac{3}{16}$. 
\begin{table}
\begin{center}
{\small 
\begin{tabular}{|c||c|c|c||c|c|c|}\hline 
$(B,J)$\,\, & $A_3$ & $B_3$ & $\beta_3$ & $A_4$ & $B_4$ & $\beta_4$ \\ 
\hline \hline 
$(0,0)$ \,\, & $-38.50$ & $(0.99 \pm 0.01)$ & $2.99 \pm 0.01$ & $-173.5$ & $ 0.184 \pm 0.007$ & $3.98 \pm 0.02$\\
$(1,1)$ \,\, & $-18.50$ & $(0.99 \pm 0.03)$ & $2.99 \pm 0.01$& $-93.25 $ &  $0.1858 \pm 0.0003$ & $ 3.97 \pm 0.02$ \\
$(0,1)$ \,\, & $-23.25 $ & $(0.99 \pm 0.02)$ & $2.99 \pm 0.01$ & $-71.5$  & $ 0.1850 \pm 0.0005$ & $ 3.96 \pm 0.02$\\
\hline
\end{tabular}
}
\caption{The parameters for the fits (\ref{fits_C3}).}\label{Table_3}
\end{center}
\end{table}
Thus, as $N_b\to\infty$, $\Lambda$ as well as $\mathcal{C}_{3,+}$ and $\mathcal{C}_{4,+}$ diverge, while the dimensionless ratios $\mathcal{C}_{3,+}/\Lambda^3$ and $\mathcal{C}_{4,+}/\Lambda^4$ approach
\begin{equation}
\lim_{\epsilon\to0^+}
\frac{\mathcal{C}_{3,+}}{\Lambda^3}\simeq 1,
\quad\quad
\lim_{\epsilon\to0^+}
\frac{\mathcal{C}_{4,+}}{\Lambda^4}\simeq -\frac{3}{16}.
\label{C3_C4_ratios}
\end{equation}

Now when $\epsilon \to 0^+$ and  $\sigma \simeq 0$, the ratio $\frac{\mathcal{C}_{4,+}}{\Lambda^4}$ reduces to 
\begin{eqnarray}
\lim_{\epsilon \to 0^+} \frac{\mathcal{C}_{4,+}}{\Lambda^4} \simeq -\frac{3}{16} \lim_{\epsilon \to 0^+}\langle \Psi_{gs,+ }| (a_1^2+a_2^2+a_3^2)^2 | \Psi_{gs,+ }\rangle
\end{eqnarray}
and hence (\ref{C3_C4_ratios}) yields 
\begin{eqnarray}
\lim_{\epsilon \to 0^+} \langle \Psi_{gs,+ }| (a_1^2+a_2^2+a_3^2)^2 | \Psi_{gs,+ }\rangle \simeq 1. 
\end{eqnarray}
This implies that 
\begin{eqnarray}
\lim_{\epsilon \to 0^+}\Big[\langle \Psi_{gs,+ }| (a_1^2+a_2^2+a_3^2)^2 | \Psi_{gs,+ }\rangle-\langle \Psi_{gs,+ }| (a_1^2+a_2^2+a_3^2)| \Psi_{gs,+ }\rangle^2\Big]\simeq 0,
\end{eqnarray}
which means the variance of $(a_1^2+a_2^2+a_3^2)$ vanishes in the limiting ground state $\displaystyle{\lim_{\epsilon \to 0^+}}  | \Psi_{gs,+ }\rangle $.

In  the singular value coordinates, $\frac{\mathcal{C}_{3,+}}{\Lambda^3}$ is given by 
\begin{eqnarray}
\frac{\mathcal{C}_{3,+}}{\Lambda^3}= 3\sqrt{3}  \langle \Psi_{gs,+ }|a_1 a_2 a_3| \Psi_{gs,+}\rangle, 
\end{eqnarray}
which, in general satisfies the expecation value arithmetic mean-geometric mean (AM-GM) inequality:  
\begin{eqnarray}
 \Big|3\sqrt{3}  \langle \Psi_{gs,+ }|a_1 a_2 a_3| \Psi_{gs,+}\rangle\Big|\leq \langle \Psi_{gs,+ }| (a_1^2+a_2^2+a_3^2)^\frac{3}{2} | \Psi_{gs,+ }\rangle. 
\end{eqnarray}
Further, as $(a_1^2+ a_2^2+ a_3^2)$ is semi-positive definite and $(a_1^2+ a_2^2+ a_3^2)^\frac{3}{2}$ is strictly convex,  the quantity on the RHS of the AM-GM inequality is bounded from below by the Jensen's inequality: 
\begin{eqnarray}
\langle \Psi_{gs,+ }| (a_1^2+a_2^2+a_3^2)^\frac{3}{2} | \Psi_{gs,+ }\rangle \geq \langle \Psi_{gs,+ }| (a_1^2+a_2^2+a_3^2) | \Psi_{gs,+ }\rangle^\frac{3}{2}. \label{Jensen_ineq}
\end{eqnarray}
In the $\epsilon \to 0^+$ limit, the vanishing variance of $(a_1^2+a_2^2+a_3^2)$ saturates the bound (\ref{Jensen_ineq}): 
\begin{eqnarray}
\lim_{\epsilon \to 0^+}\langle \Psi_{gs,+ }| (a_1^2+a_2^2+a_3^2)^\frac{3}{2} | \Psi_{gs,+ }\rangle \simeq \lim_{\epsilon \to 0^+}\langle \Psi_{gs,+ }| (a_1^2+a_2^2+a_3^2) | \Psi_{gs,+ }\rangle^\frac{3}{2}=1, \label{Jensen_ineq_sat}
\end{eqnarray}
This, along with the numerical estimate (\ref{C3_C4_ratios}), lead to satuaration of the AM-GM inequlaity
\begin{eqnarray}
  \lim_{\epsilon \to 0^+}\langle \Psi_{gs,+ }|a_1 a_2 a_3| \Psi_{gs,+}\rangle \simeq   \frac{1}{3\sqrt{3}}\lim_{\epsilon \to 0^+} \langle \Psi_{gs,+ }| (a_1^2+a_2^2+a_3^2)^\frac{3}{2} | \Psi_{gs,+ }\rangle. 
\end{eqnarray}
The above can only be satisfied if  the wavefunction $\displaystyle{ \lim_{\epsilon \to 0^+}}  | \Psi_{gs,+}\rangle $ has support purely on configurations with equal singular values. 
In other words, the saturation of the AM-GM inequality implies that the limiting ground state in the gauge configuration space localizes strictly on the surface with $a_1 \simeq a_2 \simeq a_3=\frac{1}{\sqrt{3}}$ and the probability of finding configurations with any other singular value in this state vanishes. 

It is important to note that this phenomenon of strict localization of the ground state on the surface of equal singular values does not happen for any other $g$ in the weak coupling regime: we have explicitly verified that  
\begin{eqnarray}
\lim_{\epsilon \not\to 0^+} \frac{\mathcal{C}_{3,+}}{\Lambda^3} <1, \quad\quad \lim_{\epsilon \not\to 0^+} \frac{\mathcal{C}_{4,+}}{\Lambda^4} >-\frac{3}{16}. 
\end{eqnarray}

\begin{figure}
\begin{center}
\includegraphics[width=18cm]{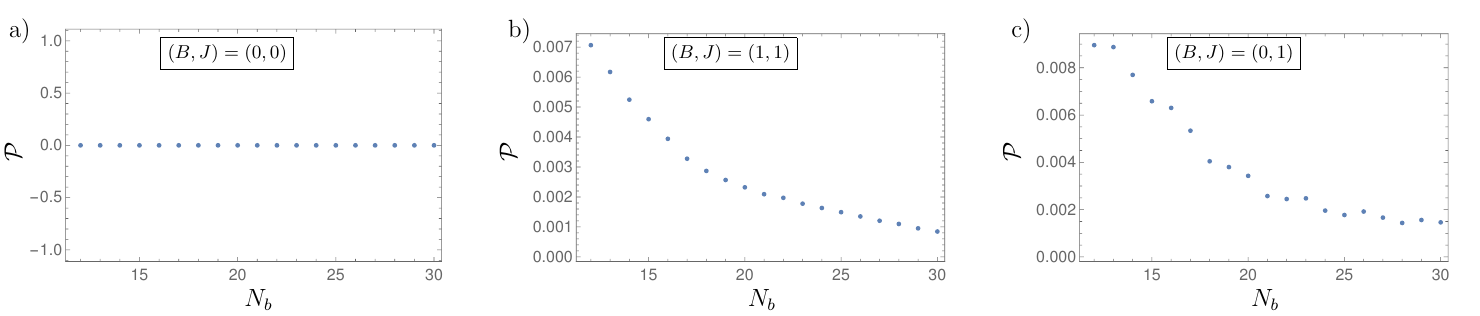}
\caption{$\mathcal{P}$ vs $N_b$ in a) the $(B,J)=(0,0)$, b) the $(B,J)=(1,1)$, and  c) the $(B,J)=(0,1)$ sectors.}\label{Fig_16}
\end{center}
\end{figure}
\textit{Spin-color locking:}  
On the surface of equal singular values, the matrix $m$ takes the form $m \simeq a Q$, where $Q= \mathcal{R}\mathcal{S}^T\in SO(3)$ and $a$  is the common singular value. Under a spatial rotation $r\in SO(3)_{rot}$,  $Q$ transforms as $Q_{ia}\to Q'_{ia} = r_{ij} Q_{ja}$. Every such rotation  can be compensated by a gauge transformation $s(r)=Q^T r Q \in SO(3)_{color}$ since $r Q s(r)^T= r Q (Q^T r^T Q)=Q$. Thus, at fixed $a$, the configurations are parametrized by $Q$, and the product group $SO(3)_{ rot}\times SO(3)_{ color}$ acts transitively on this family.  In particular, each configuration is invariant under the combined transformations $(r,s(r))$, which form its stabilizer subgroup $H_Q\equiv \left\{(r,s(r))\mid r\in SO(3)_{rot}\right\} \simeq SO(3)_{diag}$. 

This stabilizer structure imposes a constraint on the representation content of the glue states localized on this surface. To see this, consider a glue component $F(Q)$ transforming in the spin-$\ell$ representation of spatial rotations and the color representation $c$. For any $F(Q)$, we can define  $\widetilde{F}(Q)\equiv F(Q)\Big(D^{(c)}(Q)\Big)^T$ where $D^{(c)}$ denotes the $(2c+1)$-dimensional irreducible representation matrices of $SO(3)$. The invariance of $Q$ under the action of its stabilizer subgroup leads to
\begin{eqnarray}
D^{(\ell)}(r) \widetilde{F}(Q)   =\widetilde{F}(Q) D^{(c)}(r).  
\end{eqnarray}
Thus, $\widetilde F(Q)$ is an intertwining map between the color representation $c$ and the spin representation $\ell$. By Schur's lemma, such a intertwiner between two irreducible representations exists only if the representations are equivalent. Consequently, if $\widetilde{F}(Q)$, and hence $F(Q)$ is non-zero,  then $\ell =c$.  Thus localization on the surface of singular values in any state is associated with the representation-theoretic selection rule: the gluonic components of the state must carry identical irreducible representations under spatial rotations and gauge transformations. In this sense, their spin and color quantum numbers are locked.

To examine whether the state  $\displaystyle{\lim_{\epsilon \to 0^+}}| \Psi_{gs,+}\rangle$ exhibits spin-color locking and realizes this selection rule, we investigate  its gluonic representation content. In each of the three sectors, $(B,J)=(0,0),(1,1)$, and $(0,1)$, we construct the projectors $P_{\ell c}$ onto gluonic states with spin $\ell$ and color representation $c$. Defining $\mathcal{P}\equiv  \displaystyle{\lim_{\epsilon \to 0^+}}\displaystyle{\sum_{\ell \neq c}} \langle  \Psi_{gs,+} | P_{\ell c} | \Psi_{gs,+}\rangle$, we find that $\mathcal{P}$ decreases progressively with increasing $N_b$, as shown in Fig.~\ref{Fig_16}. This behavior suggests $\displaystyle{\lim_{N_b\to \infty}}\mathcal{P} \to 0$, so that the non-vanishing probability weight of $\displaystyle{\lim_{\epsilon \to 0^+}}| \Psi_{gs,+}\rangle$ is asymptotically concentrated in gluonic components with $\ell=c$. This provides a representation-theoretic signature of the spin-color locking associated with localization on the surface of equal singular values.

The spatial and gauge rotations acts on the ground state via the unitary operators $V(r)=V_{_F}(r)V_{_G}(r)$ and $U(s(r))=U_{_F}(s(r))U_{_G}(s(r))$, respectively. Here,  the subscripts $F$ and $G$ refers to the fermionic and glue sectors. The gauge invariance of the physical state implies $ U_{_F}(s(r))U_{_G}(s(r))\displaystyle{\lim_{\epsilon \to 0^+}}| \Psi_{gs,+}\rangle=\displaystyle{\lim_{\epsilon \to 0^+}}| \Psi_{gs,+}\rangle$, or equivalently $U_{_G}(s(r))\displaystyle{\lim_{\epsilon \to 0^+}}| \Psi_{gs,+}\rangle=U_{_F}(s(r)^T)\displaystyle{\lim_{\epsilon \to 0^+}}| \Psi_{gs,+}\rangle$. On the other hand, since $\displaystyle{\lim_{\epsilon \to 0^+}}| \Psi_{gs,+}\rangle$ has total spin $J$: 
\begin{eqnarray}
 V_{_F}(r)V_{_G}(r)\displaystyle{\lim_{\epsilon \to 0^+}}| \Psi_{gs,+}\rangle= D^{(J)}(r) \displaystyle{\lim_{\epsilon \to 0^+}}| \Psi_{gs,+}\rangle. 
\end{eqnarray}
Because the glue spatial rotation can be compensated by the corresponding color transformation on the surface of equal singular values, we get
\begin{eqnarray}
V_{_G}(r) \displaystyle{\lim_{\epsilon \to 0^+}}| \Psi_{gs,+}\rangle=  U_{_G}(s(r)^T)\displaystyle{\lim_{\epsilon \to 0^+}}| \Psi_{gs,+}\rangle= U_{_F}(s(r))\displaystyle{\lim_{\epsilon \to 0^+}}| \Psi_{gs,+}\rangle. 
\end{eqnarray}
This yields
\begin{eqnarray}
V_{_F}(r)U_{_F}(s(r))\displaystyle{\lim_{\epsilon \to 0^+}}| \Psi_{gs,+}\rangle= D^{(J)}(r)\displaystyle{\lim_{\epsilon \to 0^+}}| \Psi_{gs,+}\rangle. 
\end{eqnarray}
Thus, under the combined spatial rotation and its compensating color transformation, the fermionic components of the state transform as the $(2J+1)$-dimensional irreducible representation of $SO(3)$. 

We therefore find a complementary role for the glue and fermionic components in the limiting ground state $\displaystyle{\lim_{\epsilon \to 0^+}}| \Psi_{gs,+}\rangle$. The glue components become concentrated in representations with matched spatial-spin and color quantum numbers, while the fermionic components combine their spatial and color transformation properties to furnish the total angular momentum $J$.

\begin{table}[h!]
\begin{center}
 \small{
\begin{tabular}{|c|c|c|c|c|c|c|c|c|}
\hline &&&&&&\\ 
Sector & QPT & $\mathcal{I}^{BJ}_{loc}$ & $\Phi$ &  $\langle S_iS_i\rangle_{\textrm{gs}}$ &Axial charge   & quark-glue  \\
$(B,J)$ & & && &$\langle Q_0\rangle_{\textrm{gs}}$ & Interaction \\  \hline \hline 
&&&&&&\\ 
$(0,0)$ & Yes &   discontinuous & divergent &   discontinuous &  discontinuous & present\\ 
&$g_\ast^{00}\simeq 0.109$& at $g_\ast^{00}$&at $g_\ast^{00}$&at $g_\ast^{00}$&at $g_\ast^{00}$&\\ 
&&&&&&\\ 
$(1,1)$ & Yes &   discontinuous & divergent &   discontinuous &  continuous & present\\
&$g_\ast^{11}\simeq 0.118$&at $g_\ast^{11}$&at $g_\ast^{11}$&at $g_\ast^{11}$&at $g_\ast^{11}$&\\ 
&&&&&&\\ 
$(0,1)$ & Yes &   discontinuous & divergent &   continuous &  continuous & present\\
&$g_\ast^{01}\simeq 0.122$&at $g_\ast^{01}$&at $g_\ast^{01}$&at $g_\ast^{01}$&at $g_\ast^{01}$&\\ 
&&&&&&\\ 
$(2,0)$ & No QPT&   always  &always &   always  & always  & absent\\
&&continuous& finite & continuous&continuous&\\
&&&&&&\\ 
$(1,0)$ & No QPT; &   always  &always &   always  & always  & absent\\
&but a sharp&continuous& finite & continuous&continuous&\\
&crossover&&  & &&\\[1ex]
\hline
\end{tabular}
\caption{Summary of results}}
\end{center}
\end{table}
\section{Discussions}\label{results_sec}
The matrix model has proved to be a particularly successful framework for investigating the nonperturbative properties and low-energy dynamics of $SU(N)$ gauge theories. Despite being a drastic truncation of the underlying field theory, the model retains several important features of non-Abelian gauge theory, including a nontrivial gauge bundle \cite{Balachandran:2014iya, Balachandran:2014voa} and the axial anomaly \cite{Acharyya:2021egi}, and reproduces the three-color hadron spectrum with remarkable accuracy \cite{Acharyya:2016fcn, Pandey:2019dbp}. Moreover, it has revealed a rich phase structure in $SU(2)$ gauge theory coupled to fundamental or adjoint matter \cite{Pandey:2016hat, Acharyya:2017uhl, Acharyya:2024pqj, Acharyya:2026rvx}. Our results add a further dimension to this picture by identifying localization--delocalization transitions in different sectors of two-color QCD and revealing their connection with the structure of the physical Hilbert space.

 In our analysis, we have probed the localized and delocalized phases and found that their microscopic properties differ across the $(B,J)$ sectors.  It is worth emphasizing that we characterize these transitions directly in gauge-configuration space, rather than inferring them solely from spectral quantities. We also establish the essential role of the quark--glue interaction in driving these transitions. The nature of these phases at the critical point is particularly interesting. The critical behavior in gauge configuration space is characterized by $\Phi$ in the limiting ground states obtained by approaching the critical coupling from the two phases. The divergence of $\Phi$ on the delocalized side, in contrast to its finite value on the localized side, is accompanied by the emergence of distinct superselection sectors at the critical point. The two phases thus approach the critical coupling through states that become orthogonal in the limiting theory and  cannot be connected by the relevant class of physical observables.  We further show that the limiting ground state $\displaystyle\lim_{\epsilon\to0^+}|\Psi_{gs,+}\rangle$ strictly localizes on the surface of equal singular values. This is an emergent structure of the ground state: the dynamics selects a lower-dimensional submanifold within the original gauge-configuration space.  The associated spin-color locking shows that this geometric restriction has consequences for the representation content of the quantum state. To summarize,  the quark-glue interaction drives the localization–delocalization transition that not only changes the configuration-space distribution of the ground state but also its limiting quantum-mechanical and representation-theoretic structure.

In theories possessing sufficient supersymmetry, the partition function can be explicitly evaluated because the path integral localizes to a submanifold of the configuration space \cite{Pestun:2016jze}. 
This is a remarkable result. The phenomenon of localization (the critical point) observed in our matrix model seems to be a priori different because there is no SUSY in our situation.   Nevertheless, one may ask: under what conditions does the dynamics effectively reduce the relevant configuration space to a lower-dimensional submanifold?  
A satisfactory answer to this question may help generalize the ideas of localization to situations without supersymmetry. 

\appendix 

\section{Symmetries and physical Hilbert space }\label{App_symm_Hphys}
Under spatial rotations, the glue and the quark transform in spin-1 and spin-$\frac{1}{2}$ representations respectively. Their generators are
\begin{eqnarray}
L_i\equiv -2\epsilon_{ijk}  \text{Tr}\left(\Pi_j \mathcal{A}_k \right), \quad \quad S_i\equiv \frac{1}{2} (b^\dagger_{\alpha A} \sigma^i_{\alpha \beta } b_{\beta A} +d_{\alpha A} \sigma^i_{\alpha \beta } d^\dagger_{\beta A}),   
\label{spin_operator_1}
\end{eqnarray}
and satisfy $[L_i, L_j]=i\epsilon_{ijk} L_k$, and $  [S_i, S_j]=i\epsilon_{ijk} S_k. $
The Hamiltonian does not commute with $L_i$ and $S_i$ separately, but commutes with $J_i =L_i +S_i$. The $J_i$'s obey  $[J_i, J_j]=i\epsilon_{ijk} J_k$.

The glue transforms as a vector under a color $SU(2)$ rotation,  while the quark transforms in the fundamental representation. These gauge rotations in the color space are generated by 
\begin{equation}
G^a_{glue} \equiv 2i \text{Tr}\Big([\Pi_i, \mathcal{A}_i] T^a\Big), \quad \quad  G^a_{quark} \equiv   (b^\dagger_{\alpha A} T^a_{AB} b_{\alpha B} +d_{\alpha A} T^a_{AB} d^\dagger_{\alpha B})
\end{equation}
which satisfy 
\begin{eqnarray}
[G^a_{glue}, G^b_{glue}]=i \epsilon^{abc} G^c_{glue}, \quad\quad [G^a_{quark}, G^b_{quark}]=i \epsilon^{abc} G^c_{quark}. 
\end{eqnarray}
The Hamiltonian only commutes with $G^a = G^a_{glue} + G^a_{quark}$ which satisfy also $[G^a, G^b]=i \epsilon^{abc} G^c$. The operators $G^a$'s are the generators of the Gauss' law constraint.

As the fundamental representation of the gauge group is pseudo-real, the flavor symmetry (which for one flavor in 3-color QCD is the  $U(1)_V$  baryon number symmetry) in this case $SU(2)_B$ Pauli-G\"ursey symmetry. The generators of this $SU(2)_B$  are 
\begin{eqnarray}
\widehat{B}_1 &=& \frac{1}{2} \epsilon_{\alpha \beta} \epsilon_{AB} ( b^\dagger_{\alpha A} d_{\beta B} +d^\dagger_{\alpha A} b_{\beta B}),  \nonumber  \\ 
 \widehat{B}_2 &=&  \frac{i}{2} \epsilon_{\alpha \beta} \epsilon_{AB} ( d^\dagger_{\alpha A} b_{\beta B}-b^\dagger_{\alpha A} d_{\beta B} ),  \label{charge_1}  \\ 
\widehat{B}_3 &=&  \frac{1}{2} (b_{\alpha A } ^\dagger b_{\alpha A } -d_{\alpha A} ^\dagger  d_{\alpha A })\nonumber 
\end{eqnarray}
satisfy 
\begin{eqnarray}
&& [\widehat{B}_p, \widehat{B}_r] =i \epsilon_{prs} \widehat{B}_s, \quad  
[\widehat{B}_p, \widehat{Q}_0] =0, \quad [H, \widehat{B}_p] =0. 
\end{eqnarray}

The axial charge is given by $Q_0 \equiv b^\dagger_{\alpha A}b_{\alpha A} -d_{\alpha A}d^\dagger_{\alpha A}$ and when the fermions are massless, it commutes with the Hamiltonian. The axial charge generates the classical  $U(1)_A$ symmetry in the theory with massless quarks. However,  in the quantum theory this $U(1)_A$ symmetry is broken by the axial anomaly \cite{Acharyya:2021egi}.

The glue Hilbert space $\mathcal{H}_G=\displaystyle{L^2(M_3(\mathbb{R}),\prod_{ia} dM_{ia})}$ is infinite dimensional and the glue states may be organized in representations of spin and color. These glue states can have integer spin and transforms in odd-dimensional irreps of color $SU(2)$.  The quark Hilbert space $\mathcal{H}_F$ is  256-dimensional, and the fermionic states can have integer or half-integer spin $0\leq s \leq 2$ and   transform in (2r+1)-dimensional representations of color $SU(2)$ with $0\leq r\leq 2$. 
The total Hilbert space $\mathcal{H}$ is the direct product space $\mathcal{H}_F \otimes \mathcal{H}_G$.  
The Gauss law constraint demands that all physical observables  commute with $G^a$. 
The physical Hilbert space $\mathcal{H}_{phys}$ contains vectors which are annihilated by the Gauss generators: $G^a | \Psi \rangle =0$ for all $| \Psi \rangle \in \mathcal{H}_{phys}$ \cite{Pandey:2019dbp}.   
The composite states in $\mathcal{H}$ can only be colorless, when the quark states transform in odd-dimensional irreps of color $SU(2)$. This is only possible when the states have even number of quarks or anti-quarks. Thus, the physical states in $\mathcal{H}_{phys}$ always have even number of fermions.

 $\mathcal{H}_{phys} $ can be spanned by colorless eigenstates of the Hamiltonian. Because $H$ commutes with $J_i$  and $\widehat{B}_p$, the color-singlet energy eigenstates can also chosen to be  simultaneous eigenstates of $(J_iJ_i)$,  $J_3$, $(\widehat{B}_p \widehat{B}_p)$ and $\widehat{B}_3$.  Such a color-singlet energy eigenstate  
 is  labelled by the quantum numbers $\{J,J_3, B, B_3\}$ where $J(J+1)$ and $B(B+1)$ are the eigenvalues of  $(J_iJ_i)$ and $(\widehat{B}_p \widehat{B}_p)$, respectively. 
The colorless energy eigenstates necessarily have an even number of quarks: a state with an odd number of quarks carry half-integer color-charge, which cannot combined with any glue state to produce a color-singlet state.  
The fermionic Hilbert space is finite dimensional and  it is easy to see that the physical states can have max$(B)=2$.  

The glue states and the states with even number of quarks  transform in the integer representation of the algebra generated by $L_i$ and $S_i$, respectively.  Consequently, the colorless eigenstates of the Hamiltonian always transform in an integer representation of  $SO(3)_{rot}$: $J=0,1,2 \ldots$  
Further, states with even number of fermions can be of three distinct types: i) when $B_3=0$, the states are mesons with equal number of quarks and anti-quarks, ii) when $B_3= \pm1$, the states are diquarks/anti-diquarks with  the number of quarks and anti-quarks differing by two, and iii) when $B_3= \pm 2$, the states are tetraquarks/anti-tetraquarks with number of quarks  and anti-quarks  differing by four.

{
\section{$B_3$ and $J_3$ independence of $\mathcal{I}^{11}$}\label{J3B3_ind_I11}
$SU(2)_B$ acts only on the fermionic Hilbert space and leaves the gauge configuration $M_{ia}$ unchanged. Consequently, 
a generic $SU(2)_B$ rotation mixes the states with different $B_3$ but the fermionic norm $\rho^{J_3}_{_\mathcal{G}}(M)$ for any fixed $J_3$ remains invariant.  Hence, for fixed $J_3$, the value of integral $\mathcal{I}^{11}$ is independent of $B_3$. 
 
 The $J_3$-independence of $\mathcal{I}^{11}$ is particularly transparent in the Cartesian basis of the spin-1 triplet: $\{|\Psi_{\mathrm{gs}}^{i, B_3}\rangle: \, i=1,2,3\}$. The states in the spherical and cartesian basis are related by a unitary transformation: $|\Psi_{\mathrm{gs}}^{i,B_3}\rangle=U_{i,J_3}|\Psi_{\mathrm{gs}}^{ J_3, B_3}\rangle$, with $\displaystyle{\sum_{J_3}}U^{\ast}_{j,J_3} U_{ i,J_3}=\delta_{ij}$. Defining
\begin{eqnarray}
\mathcal{K}_{ij} \equiv \int_\Gamma\langle\Psi_{\mathrm{gs}}^{i, B_3}|\Psi_{\mathrm{gs}}^{j, B_3}\rangle_{_F}  \,d\mu_M, 
\end{eqnarray}
it is easy to see that for a fixed $J_3$
\begin{eqnarray}
\mathcal{I}^{11} = U^{\ast}_{i,J_3} U_{j,J_3} \mathcal{K}_{ij}.  \label{IK_rel}
\end{eqnarray}

The proper cubic rotation group  $\mathcal{C}$ leaves the region $\Gamma$ and $d\mu_M$ invariant. Under a cubic rotation $r \in \mathcal{C}$, the matrix $M$ and the state $|\Psi_{\mathrm{gs}}^{i, B_3}\rangle$  transforms as
\begin{eqnarray}
M \to r^{-1} M, \quad\quad |\Psi_{\mathrm{gs}}^{i, B_3}(M)\rangle \to r_{ij} |\Psi_{\mathrm{gs}}^{j, B_3}(r^{-1} M)\rangle. \label{M_psi_trans}
\end{eqnarray}
From above, it is easy to see that $ \mathcal{K}_{ij}$ transforms as rank-2 tensor: 
\begin{eqnarray}
 \mathcal{K}_{ij} \to  r_{ik} r_{jk'}  \mathcal{K}_{kk'}. \label{K-trans}
\end{eqnarray}

We can define a coordinate transformation  as $M'=r^{-1} M$ with $r \in \mathcal{C}$ and rewrite $ \mathcal{K}_{ij}$ as 
\begin{eqnarray}
 \mathcal{K}_{ij}&=&    \int_{\Gamma}  \langle\Psi_{\mathrm{gs}}^{i, B_3}(rM')|\Psi_{\mathrm{gs}}^{j, B_3}( rM')\rangle_{_F}  \,d\mu_{M}.
\end{eqnarray}
Using (\ref{M_psi_trans}) and the invariance of the measure $\,d\mu_{M}=\,d\mu_{M'}$, we obtain
 \begin{eqnarray}
 \mathcal{K}_{ij}&=& r_{ik} r_{jk'}  \int_{\Gamma}  \langle\Psi_{\mathrm{gs}}^{k, B_3}(M')|\Psi_{\mathrm{gs}}^{k', B_3}( M')\rangle_{_F}  \,d\mu_{M'}=  r_{ik} r_{jk'} \mathcal{K}_{kk'}. \label{K_inv}
\end{eqnarray}
Combining (\ref{K-trans}) and (\ref{K_inv}), we find that $K_{ij}$ remains invariant under $\mathcal{C}$. As the only rank-2 tensor which can remain invariant is proportional to the identity, we get 
\begin{eqnarray}
\mathcal{K}_{ij} = K \delta_{ij}. 
\end{eqnarray}

Inserting this in (\ref{IK_rel}), we obtain
\begin{eqnarray}
\mathcal{I}^{11} = K U^{\ast}_{i,J_3} U_{j,J_3} \delta_{ij}= K  U^{\ast}_{i,J_3} U_{i,J_3} = K.
\end{eqnarray}
Therefore, the localization integral is independent of $J_3$. In fact, the result holds for any normalized linear combination of the three states in the $J=1$ ground-state manifold.

The same arguments can be used in the $(B,J)=(0,1)$ sector and $\mathcal{I}^{01}$ is independent of $J_3$ as well. 
}

\end{document}